\documentclass[11pt]{article}

\usepackage{graphicx}
\usepackage{amsmath}
\usepackage{subfigure}
\usepackage{psfrag}

\newcommand{\inputFigure}[2]{\includegraphics[#1]{#2}}

\numberwithin{equation}{section} 	

\newcommand{\FullGF}{T(\vec{x},\vec{x}_{i})}
\newcommand{\FreeGF}{G_0(\vec{x},\vec{x}_{f})}
\newcommand{\GeneralGF}{F(\vec{x},\vec{x}_{f})}
\newcommand{\DotDNormal}{\,.\,d\/\vec{n}}
\newcommand{\DotProd}{\,.\,}
\newcommand{\LSquared}{$\mathcal{L}^{2}$\ }
\renewcommand{\vec}[1]{\boldsymbol{#1}}

\usepackage{amsfonts}
\newcommand{\field}[1]{\mathbb{#1}}
\DeclareMathOperator{\Tr}{Tr}
\DeclareMathOperator{\Imag}{Im}
\DeclareMathOperator{\Real}{Re}
\renewcommand{\Im}{\Imag}		
\renewcommand{\Re}{\Real}		

\begin{document}

\title{
       Exploring Periodic Orbit Expansions and Renormalisation\\
       with the Quantum Triangular Billiard.
}
\author{
        Carmelo Pisani\thanks{email:\tt pisani@maths.mu.oz.au}\\
	Department of Mathematics\\
	University of Melbourne\\
	Parkville 3052\\
	Victoria, Australia
}
\date{25 January 1996}
\maketitle
\begin{abstract}
A study of the quantum triangular billiard requires consideration of 
a boundary value problem for the Green's function of the Laplacian on 
a trianglar domain.  Our main result is a reformulation of this 
problem in terms of coupled non--singular integral equations.  A 
non--singular formulation, via Fredholm's theory, guarantees 
uniqueness and provides a mathematically firm foundation for both 
numerical and analytic studies.
We compare and contrast our reformulation, based on the exact solution for
the wedge, with the standard singular integral equations using 
numerical discretisation techniques.  We consider in detail the
(integrable) equilateral triangle and the Pythagorean 3-4-5 triangle.  
Our non--singular formulation produces results which are well behaved
mathematically.  In contrast, while resolving the eigenvalues very well,
the standard approach displays various behaviours demonstrating the need
for some sort of ``renormalisation''. 
The non-singular formulation provides a mathematically firm basis for the
generation and analysis of periodic orbit expansions.  We discuss their
convergence paying particular emphasis to the computational effort required
in comparision with Einstein--Brillouin--Keller quantisation and
the standard discretisation, which is analogous to the
method of Bogomolny.  We also discuss the generalisation of our 
technique to smooth, chaotic billiards.
\end{abstract}


\section{Introduction}
\label{sec:intro}


The main focus of this paper is on contemporary issues in quantum 
chaos. In particular the convergence problems of periodic orbit 
expansions, the computational effort needed in their generation, and 
their status as the generalisation of Einstein--Brillouin--Keller 
quantisation to classically chaotic systems. To begin we 
consider the place our problem has in a broader context.

\subsection{The need for a semiclassical limit.}

Many problems in mathematical physics reduce to the solution of 
boundary value problems for partial differential equations or infinite 
matrices: wave propagation, diffusion, lattice statistical mechanics, 
stochastic problems, thermodynamic formalism in classical 
mechanics, an so on. Often, the ``observables'' can be expressed in terms 
of the eigenvalues and eigenvectors of some operator.  The spectral 
problem of operators is of wide interest in mathematics as well. 
 
Unfortunately, these problems can only be solved in integrable cases, in
which an $n$--degree of freedom problem reduces to $n$ 1--degree of freedom
problems. By ``solve'' we mean an effective, efficient, mathematically
well behaved means of calculation which affords an opportunity for
intuitive understanding.  The only way to proceed is to recognise that the
problem has to be divided in two: the lower energy, long wavelength,
portion of the spectrum and the high energy, short wavelength, portition of
the spectrum.  The former regime is approximately reduced to the
diagonalisation of a finite matrix.  In the latter regime one uses
semiclassical approximations.  The stategy then is to develop the two
techniques sufficiently and piece together their results to obtain a
\emph{global} understanding of the spectrum.

For practical purposes the only global information one often needs is
Weyl's density of states. Other times one simply requires the ground and
first few excited states.  It is then important to know at what point the
spectrum crosses over into the ``semiclassical'' regime. It appears that
the semiclassical regime extends quite far down.  In the case of the disk,
using asymptotic approximations for the Bessel function, one finds that a
semiclassical approach gives acceptable approximations even for the ground
state.  For special examples such as the equilateral triangle and other
problems solved by ``Bethe Ansatz'' the semiclassical approach gives the
exact answer. For cases where global, detailed, spectral information is
needed semiclassical methods are indispensible. 

\subsection{Semiclassical techniques}

For historical reasons much of the development of semiclassical 
techniques has been driven by practical issues in optics and quantum 
mechanics.  Such motivations have lead to mathematical developments 
such as the WKB method.  Most of our motivation will come from quantum 
mechanics but optics will play a role.

Semiclassical techniques have long had an important role in quantum 
mechanics.  In the early days of the old quantum theory (and in fact 
in some contemporary problems) they provided the \emph{only} 
quantisation technique.  Failing to quantise more complicated systems, 
notably the helium atom, they were replaced by quantum mechanics.  
There, through the WKB method, they reappeared as approximation 
techniques in 1--dimensional or separable problems.

Multidimensional, non--integrable quantum problems are much more difficult
and are usually treated with integral equations or diagonalisation in some
basis.  While providing theoretical results to compare with experiment,
such methods are computationally intense and afford little opportunity for
intuitive understanding.  Semiclassical techniques were first widely
applied to such problems in optics~\cite{KellerGTD62}.  They have also
been applied to quantum chemistry~\cite{MillerRev} and more general
problems of partial differential equations~\cite{KellerSiam}.

\subsection{Semiclassical quantisation.}

Although it was not widely realised, prior to the late 60s semiclassical
quantisation was restricted to classically integrable systems using
Einstein--Brillouin--Keller (EBK)
quantisation~\cite{GutzwillerBook,KellerSiam,SteinerRev94}.  It was then
that Gutzwiller~\cite{GutzwillerOrig} developed a method of semiclassical
quantisation based on the calculation of the Green's function via a Laplace
transform of the time--dependent propagator (heat kernel).  For the latter 
a semiclassical approximation could be derived via Feynman's path integral
regardless of the integrablity of the corresponding classical system.
Gutzwiller used the resulting periodic orbit expansion to determine the
spectrum of the Kepler problem, and later, an anisotropic version of it. 

Around the same time, Balian and Bloch~\cite{BalianBloch} obtained a
similar result while extending a classical result of Weyl for the density
of states of the Laplacian. They used a multiple expansion technique for
the Green's function due originally to Carlemann.  Weyl's formula is only
relevant if one wishes to calculate certain sums over the spectrum.  Balian
and Bloch were interested in refining the Weyl result to incorporate the
``shell structure'' of the spectrum. They appear to have been motivated by
certain spectral sums arising in nuclear physics.

In recent years, mostly driven by resurgent interest in classical 
chaos, there has been renewed interest in this work.  The prime 
motivation has been to rehabilitate the old quantum theory by 
providing a quantization rule, analogous to EBK quantisation, for 
non-integrable systems.  It is well known~\cite{GutzwillerBook} that 
periodic orbit expansions are absolutely divergent for chaotic systems 
due to an exponential proliferation of periodic orbits.  Despite this, 
they have had some empirical success as rules for ``quantising 
chaos''.  This motivatives the oft--stated hope that the series are 
conditionally convergent.  The well--known properties of conditionally 
convergent series with regards rearrangment~\cite{SeriesRef} then begs 
the question: In what order should the sum be carried out?

In an effort to overcome this difficulty, the original periodic orbit
expansions have been refined in many ways.  These refinements tend to be
motivated from results in other areas of mathematics and physics: the
theory of the Riemann zeta function, topology of hyperbolic flows in
classical chaos, the boundary integral method for partial differential
equations, and scattering theory.  These refinements include: the
Riemann--Siegel lookalike formula and pseudo--orbit
expansions~\cite{BerryKeating}, cycle expansions~\cite{Cvitanovic}, surface
of section techniques \cite{Bogomolny} and inside--outside duality
\cite{Smilansky}.

Some refinement of Weyl's formulae in the calculation of spectral sums has
been motivated from the Selberg trace formula for geodesic motion on
hyperbolic manifolds~\cite{SteinerRev94,SteinerSmoothed}.

These techniques have been tested on a number of chaotic systems 
including: the hyperbola billiard \cite{SieberPhD}, the helium atom 
\cite{HeliumAtom}, the wedge billiard in a gravitational field
\cite{TzerdiGoodings93,SmilanskyRouvinez95}, the 
Sinai billiard \cite{BerrySinai,SmilanskySinai}, the 3 disk system 
\cite{Cvitanovic,Wirzba92}, and smooth Hamiltonian 
systems \cite{Haggerty}. Much progress has been made, as can most readily 
be seen in \cite{SieberPhD} and \cite{TzerdiGoodings93} where various 
modifications are compared.

Despite the relative success of these improvements, they have all come 
from well motivated, but largely \emph{heuristic} arguments and 
derivations.  The well known mathematical difficulties are not 
directly tackled.  An exception to this is some recent and very 
important work of Georgeot and Prange~\cite{GeorgeotPrange95}.  
Starting with the integral equations of Balian and Bloch they apply 
the Fredholm theory of integral equations~\cite{Smithies62} to show, 
amongst other things, that the pseudo--orbit expansions for smooth 
billiards \emph{converge}.  The proper 
order of summation, of what is clearly a conditionally convergent 
series, is dictated by the Fredholm theory.  Thus, at least for smooth 
billiards, it seems that the long standing questions regarding 
convergence have been settled. (well, maybe.  See 
section~\ref{sec:discussSmoothBill}.)

Apart from convergence issues, there is the equally serious question 
of computational effort.  It is fairly clear from the literature that, 
unlike EBK quantisation, periodic orbit expansions become more and 
more computationally intensive as one seeks higher and higher 
eigenvalues.  Furthermore this remains the case even with refinements 
which improve convergence~\cite{CvitanovicChaos92}.  While it is clear 
that exponential proliferation is the cause, it contradicts 
the correspondance principle which suggests that the semiclassical 
approximation should improve with ``larger quantum numbers''. It may 
be, however, that this expectation is intertwined with our ``integrable'' 
intuition and needs to be reassessed in these ``chaotic'' times.

Thus, despite the progress made, mystery continues to surround 
semiclassical quantisation of non--integrable systems.

\subsection{A role for the triangular billiard}

Our aim is to critically examine of the nature of semiclassical 
quantisation rules.  To do this we seek out a model problem which is 
sufficiently simple to enable a first principles derivation
which maintains reasonable standards of mathematical 
rigour.  Whereas we do not seek, and are not capable of, a rigorous 
derivation, we aim at one which can plausibly form the basis for a 
rigorous approach.  Our choice is the {\em quantum triangular 
billiard}.

The quantum triangular billiard is well suited to a first-principles 
derivation.  Being a billiard, the boundary value problem for the 
Green's function can be converted to an integral equation.  This 
permits the application of a substantial body of standard theory.  As 
with Georgeot and Prange, our main tool will be the Fredholm theory of 
integral equations.  Most importantly, piecewise linear boundaries 
afford us much needed flexibility. Our discussion is immediately applicable
to polygonal billiards.  Although our notation will have an eye toward this
generalisation, we will limit ourselves to triangles so as not be be
distracted from our aim.  The prospects for handling more general
billiards, in particular those whose boundaries have smooth components,
will be discussed in Section~\ref{sec:discussSmoothBill}.

The quantum triangular billiard has been studied in a number of 
contexts.  In the context of quantum chaology it has not attracted 
much attention because polygonal billards are not 
chaotic~\cite{Gutkin86} (see however~\cite{FordPolygon93}) so that one 
does not have exponential proliferation of periodic orbits.  Despite 
this, it is still (generically) a non-integrable system, 
intermediate between integrable systems and chaotic ones, and so still 
of interest in its own right.
It has also been considered in the contexts of eigenvalue 
degeneracies~\cite{BerryWilkinson}, energy level 
statistics~\cite{MiltenbergRuijgrok}, and modifications of periodic 
orbit expansions due to non-isolated and diffractive periodic 
orbits~\cite{PavloffSchmit95}.  Integrable triangles have 
eigenfunctions which are exactly expressed in terms of a 
finite superposition of plane waves via the Bethe 
Ansatz~\cite{Mattis}.  The general triangle has also been studied in 
an effort to extend the Bethe ansatz~\cite{Gaudin}.  The Weyl formula 
for polygonal billiards has long been known~\cite{KacDrum66}.  More 
recently this result has been extended to calculate determinant of the 
Laplacian~\cite{AurellSalomonson94}.

\subsection{Statement of problem}

To be more precise, the task we have set ourselves is the calculation of the
spectrum of the Dirichlet Laplacian on a triangular domain. All spectral
information conveniently resides in the \emph{Green's function} which is the
unique solution of the boundary value problem
%
%
\begin{subequations}\label{eq:TriangleBVP}
\begin{gather}
  \nabla^{2}\FullGF + z\,\FullGF 
  = 0,
  \quad\mathbf{x}\in\triangle/\{\mathbf{x}_i\} 
  \label{eq:helmholtzT}  
  \\
  \lim_{\epsilon \to 0} \oint_{C_{\mathbf{x}_i}}
  \nabla\FullGF\DotDNormal
   =  1 
  \label{eq:helmholtzTsource} 
  \\  
  \FullGF 
   =  0 
  \quad \text{ for }\mathbf{x}\in\partial\triangle 
  \label{eq:helmholtzTbc}
\end{gather}
\end{subequations}
$C_{\mathbf{x}_i}$ is a circular contour of radius $\epsilon$ taken
counterclockwise around the ``source'' point, $\mathbf{x}_i$. $\triangle$
denotes the triangular domain and $\partial\triangle$ denotes its boundary.

The unit source condition, (\ref{eq:helmholtzTsource}), is usually specified
as a Dirac $\delta$--function inhomogeneity in (\ref{eq:helmholtzT}). We
take the opportunity to remind the reader of this more precise description,
even if it is somewhat old fashioned~\cite{SommerfeldPdes}.

We are now in a position to be precise about the semiclassical limit. In
this problem there are two natural length scales: the ``de Broglie''
wavelength $1/\sqrt{z}$ and any length $L$ whose order of magnitude
reflects the geometric distances over which the domain varies. For
$\triangle$ it is the length of the sides. More complicated domains may
provide other length scales which may or may not be a similar order of
magnitude.  The semiclassical limit is where the ``extent'' of the triangle
is much larger than the ``de Broglie'' wavelength, $\sqrt{z} L \gg 1$.  
Cast in this mathematical language, the semiclassical limit becomes a
problem of asymptotic analysis.

\subsection{Standard integral equation approach}

For generic triangles the classical problem is non--integrable so that EBK
is not applicable.  The only way to obtain further analytic insight is to
convert the boundary value problem into an integral equation using Greens'
theorem.  This is a standard technique in mathematical physics with a long
history.  It is widely used in electrostatics~\cite{SommerfeldPdes},
diffraction~\cite{BornWolf80}, and quantum scattering
theory~\cite{Joachain79}, both as a way to derive integral equations and
also as a starting point for analytic and numerical approximations.  It's
(indirect) use in the study of quantum billiards began with Balian and
Bloch~\cite{BalianBloch}.  We shall refer to this technique as the
``standard approach'' or the ``free--based method''.

For the Helmholtz equation various types of integral equations can be 
obtained.  For the eigenfunctions one obtains an homogenous integral 
equation of the second kind.  For the Greens' function one obtains an 
inhomogenous integral equation the second kind.  Whilst there is a 
close connection between the two, the Fredholm theory gives the 
results needed to generate periodic orbit expansions only in the 
inhomogenous case.  One is thus forced to study the spectral problem 
via the Green's Function.

It is well known~\cite{Riddell} (though sometimes overlooked) that in
contrast with smooth billiards, the standard integral equations for
polygonal billiards are \emph{singular}. By this we mean that the
requirements on the kernel demanded by Fredholm theory are not satisfied.
Singular integral equations are known to be pathological. For example, the
Lippmann--Schwinger equations for the quantum three body problem, which are
singular, do not have unique solutions~\cite{Joachain79}. In the literature on
the Helmholtz equation, singular equations are known to yield
``spurious solutions''~\cite{BerryWilkinson,Riddell}. Although useful
information can be extracted from these equations, the presence of
pathologies creates the need for \emph{ad--hoc} techniques,
caution, and ultimately, \emph{doubt}.

The main result of this paper is the derivation of \emph{non--singular}
integral equations for the Green's function of the quantum triangular
billiard. The derivation of non--singular integral equations from
singular ones has a precedent in the work of Faddeev on the quantum three
body problem~\cite{Joachain79}. The basic idea of Faddeev's work is that it is
neccessary to treat part of the interaction in an explicit non--perturbative
manner. Faddeev does this by eliminating the potential in favour of the
T--matrix. As we will discuss in Section~\ref{sec:DirectFaddeev} a direct
application of Faddeev's technique is not convenient for our problem.
Instead we follow the spirit by basing a derivation on the Green's function
for the wedge. 

Whereas most treatments of the quantum billiard problem depict the
derivation of the integral equations as straightforward and concentrate on
the generation of periodic orbit expansions, we shall show that for
polygonal billiards this is not the case. Although we shall present a
relatively compact and well motivated derivation, this does not reflect the
tortuous path we took to reach it.

\subsection{Outline of paper}

We begin in Section~\ref{sec:FreeBased} with an outline of the standard
approach, presenting more detail than is customary in order to demonstrate
some subtleties which are usually overlooked. The derivation will be
accompanied by an elaboration of the relatively well known intepretation in
terms of a ``quantum Poincar\'e section''. We then demonstrate the singular
nature of the standard integral equations and discuss ways in which they
can be manipulated to obtain non--singular equations. We also emphasise the
perturbative nature of the whole Green's theorem--based approach. We then
warm up for the wedge--based approach by rederiving the standard integral
equations from a perturbation about the half--plane Green's function. A
similar derivation has recently been presented by Li and
Robnik~\cite{LiRobnik95}.

In Section~\ref{sec:WedgeBased} we outline the derivation of 
non--singular integral equations based on the exact solution for 
the wedge Green's function.  The implementation of the non--singular 
formulation requires a detailed analytic and computational 
understanding of this solution.  In Section~\ref{sec:Kernel} we give 
the necessary technical detail for the wedge--based kernel.  In the 
appendix we give an outline of the derivation of the wedge Green's 
function.  In Section~\ref{sec:Numerics} we present and compare, for 
purposes of validation, the results obtained from a standard numerical 
treatment of both the singular and non--singular integral equations.  
In particular we present the eigenvalues for the equilateral and 
Pythagorean 3-4-5 triangles by examining the zeros of the Fredholm
determinant. We show that the Fredholm determinant for the discretised
singular equation does not converge although it resolves the eigenvalues
well. In contrast, the non--singular equation behaves well, as the Fredholm
theory demands.

Given a suitable integral equation, the derivation of periodic orbit 
expansions is more or less mapped out by Georgeot and Prange.  Let 
us briefly outline their discussion. The Fredholm theory expresses 
the solution of an integral equation in terms of a resolvent kernel 
$H_{\lambda}$ and a Fredholm determinant $\Delta(\lambda)$ in the form
\begin{equation}
x(t) = x_0(t) + \frac{1}{\Delta(\lambda)}
				\int_{a}^{b}H_{\lambda}(t,s)
                  x_0(s)\, ds
\end{equation}
\cite{Smithies62}. The eigenvalues are then the zeros of the Fredholm 
determinant, which one identifies as a zeta function.  The Fredholm 
determinant can be expanded in series using the identity $\log \det ( 
I - K) = \Tr \log (1 - K)$ \cite{AlonsoGaspard93}.  The expansion of 
$\log ( I - K )$ in powers of $K$ has a finite radius of convergence.  
However exponentiating the series and re--expanding gives the standard 
Fredholm series which has an infinite radius of 
convergence~\cite[Theorem 6.5.2]{Smithies62}.  A semiclassical 
expansion of the traces $\Tr(K^{n})$, which appear in this series, then 
provides a \emph{convergent} periodic orbit expansion for this zeta 
function which one identifies with the psuedo--orbit expansion.

At this point it is necessary to make a terminological aside.  The term 
``periodic orbit expansion'' can refer to the original Gutzwiller 
series or any of its refinements.
The context of discussion allows one to use it without 
qualification.  However, in order to emphasise the firm foundation 
provided by the Fredholm theory, we feel that it is perhaps appropriate 
to occasionally use the term ``semiclassical Fredholm series''.

In Section~\ref{sec:gtf} we use the Fredholm theory to analyse the 
convergence of resulting periodic orbit expansion.  Here it will 
become clear that while Fredholm theory solves the convergence problem 
the number of terms needed for convergence grows like $\sqrt{E}$.  The 
semiclassical approximation of the $n$--th term in the series expansion of 
the Fredholm determinant requires all $n$--bounce periodic orbits.  
For the triangle the number of such orbits grows polynomially in n.  
The computation effort in the periodic orbit expansion is then 
polynomial in $\sqrt{E}$.  In contrast the effort for chaotic systems 
is exponential.  We compare this effort with the $O(n^{3})$ effort needed 
in the numerical discretisation of the integral equations. This 
comparision clearly defines how far we are from the goal of emulating 
EBK quantisation for chaotic systems.

We conclude with a discussion in Section~\ref{sec:discuss}. Here we present
the singular equation as a simple model system which requires
``renormalisation''.  We note the presence of logarithmic singularities in
the Fredholm series making it comparable to perturbation series in quantum
field theory. The wedge based theory is then interpreted as a first
principles, mathematically firm, ``renormalisation scheme'' where any
\emph{ad--hoc}``regularisation'' and ``subtraction of infinities'' is
absent. We then present arguments that cast doubt on the validity of the
Balian--Bloch integral equations for smooth billiards.  We then suggest
ways in which our construction can be carried over \emph{in principle} to
smooth billiards.  We also discuss the relationship between the Fredholm
series approach and other modifications of Gutzwiller's original trace
formula.  We also present our construction as a way to derive, in a
mathematically satisfactory way, Keller's geometrical theory of
diffraction~\cite{KellerGTD62}.

\section{Standard Approach: Singular Integral Equations}
\label{sec:FreeBased}
\subsection{Standard pre--integral equation}
\label{sec:PreIntEqn}

The standard approach is based on the free Greens function. This is the
solution of the Helmholtz equation on the Euclidean plane which decreases at
infinity and has a unit source.
%
%
\begin{equation}
  \FreeGF =
  -\frac{i}{4}H_{0}^{(1)}(\sqrt{z}|\vec{x}-\vec{x}_{f}|)
\end{equation}
%
%
where $H_{0}^{(1)}(z)$ is the zeroth order Hankel function of the 
first kind~\cite{Abramowitz68}.

Our main point is that this is a \emph{choice} and that for
polygonal billiards it is a poor choice. Let us thus begin with \emph{any}
Greens' function, i.e. any function $\GeneralGF$ which is a solution of the
Helmholtz equation with a unit source. The domain on which $\GeneralGF$ is
defined and the boundary conditions it satisfies are totally arbitrary.

\begin{figure}[htb]
    \centering
    \psfrag{P0}[][]{\Large $\vec{P}_0$}
    \psfrag{P1}[][]{\Large $\vec{P}_1$}
    \psfrag{P2}[][]{\Large $\vec{P}_2$}
    \psfrag{r0}[][]{$\mathbf{r}_0(t)$}
    \psfrag{r1}[][]{$\mathbf{r}_1(t)$}
    \psfrag{r2}[][]{$\mathbf{r}_2(t)$}
    \psfrag{n0}[][]{$\mathbf{n}_0$}
    \psfrag{n1}[][]{$\mathbf{n}_1$}
    \psfrag{n2}[][]{$\mathbf{n}_2$}
    \psfrag{xi}[][]{$\mathbf{x}_i$}
    \psfrag{xf}[][]{$\mathbf{x}_f$}
    \psfrag{x(s)}[][]{$\mathbf{x}(s)$}
    \psfrag{t}[][]{$t$}
    \psfrag{d}[][]{$\delta$}
    \inputFigure{height=0.3\textheight}{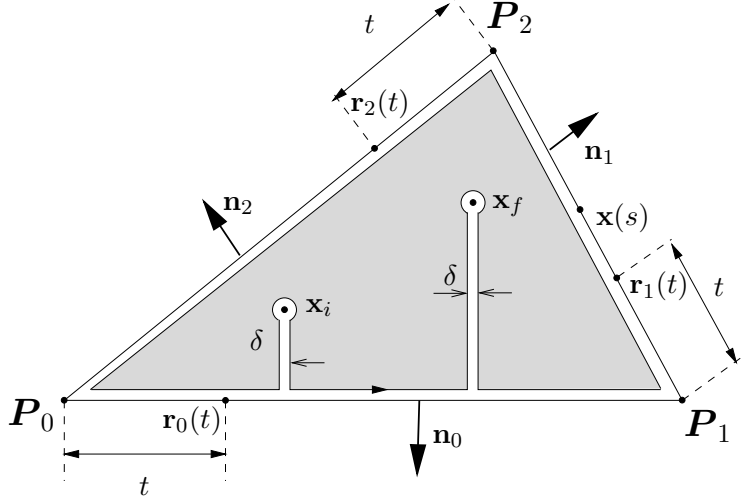}
    \caption{Contour for derivation of pre--integral equation. 
    For convenience distances $t$ are
    defined with respect the nearest vertex in a counterclockwise sense.
    The heads of vectors are denoted using points. The integration region
    $D$ is shaded and the sense of $\partial D$ is shown.
  }
    \protect\label{fig:IntPath}
\end{figure}

Applying Green's second identity,
\begin{equation}
  \int_{D} (\phi\nabla^2\psi - \psi\nabla^2\phi)\,dx\,dy = 
  \int_{\partial D} (\phi\nabla\psi - \psi\nabla\phi)\DotDNormal,
\end{equation}
with $\phi = \FullGF$ and $\psi = \GeneralGF$ to the domain $D$ given in
Figure~\ref{fig:IntPath}, taking the width of the
channels $\delta$ to zero, and using continuity in the
usual way, one obtains
%
%
\begin{equation} \label{eq:AfterApplyingGreensIdentity}
  \left(\int_{\partial\triangle}
    - \int_{C_{{\vec{x}_{i}}}} 
    - \int_{C_{{\vec{x}_{f}}}}
  \right)
  [\FullGF \nabla \GeneralGF - \GeneralGF \nabla \FullGF
  ]\DotDNormal = 0.
\end{equation}
%
%
$C_{\vec{x}_{i}}$ and $C_{\vec{x}_{f}}$ are circular contours of
radius $\epsilon$. 

In the integral over $C_{\vec{x}_{i}}$ all functions, except $\nabla 
T$, are continuous and can be approximated to $O(\epsilon)$ by their 
values at $\vec{x}_{i}$.  Using this, the unit source condition, and 
taking $\epsilon \to 0$ we obtain
%
%
\begin{equation}
\begin{split}
  \lim_{\epsilon \to 0} \oint_{C_{\vec{x}_{i}}}
  &
  \left[\FullGF \nabla \GeneralGF 
  - \GeneralGF \nabla \FullGF
  \right] \DotDNormal
  \\
  & =
  -F(\vec{x}_i, \vec{x}_f)
  \left[
    \lim_{\epsilon\to 0} \int_{C_{\vec{x}_{i}}}
    \nabla \FullGF \DotDNormal
  \right]
  \\
  &  =
  - F(\vec{x}_i, \vec{x}_f)
\end{split}
\end{equation}
%
%
The integral over $C_{\vec{x}_f}$ is evaluated similarly.

Using the Dirchlet boundary condition satisfied by $\FullGF$ we finally
obtain
\begin{equation}
  \mathrm{T}(\vec{x_{f}},\vec{x_{i}}) = 
  \mathrm{F}(\vec{x_{i}},\vec{x_{f}}) -
  \int_{\partial\triangle}
  \GeneralGF \nabla \FullGF 
  \DotDNormal       
  \label{eq:PreIntegralEquation}
\end{equation}
The integration in \eqref{eq:PreIntegralEquation} is concentrated on the boundary
of the domain but also involves values of the unknown function at arbitrary
points in the interior. Interpreted as an integral equation on 
$\triangle$ it has a very
singular kernel. The standard approach instead regards this as an equation
which determines the solution in the interior \emph{given} it's normal
gradient on the boundary. For this reason we refer to
\eqref{eq:PreIntegralEquation} as the \emph{pre--integral equation}.
Let us pause for a moment to interpret this equation.

\subsection{Poincar\'e section interpretation}

The boundary of the domain in \eqref{eq:PreIntegralEquation} plays 
a distiguished role very similar to its role in the classical problem.  
In the classical problem Birkhoff defined a Poincar\'e section for 
billiards consisting of pairs $(q,p)$ giving the position and 
tangential momentum at each collision with the boundary.  A particular 
classical orbit is then represented in the Poincar\'e section by a 
sequence of points, $(q_n,p_n), n=0,1,2\ldots,\infty$.  Knowing that 
the motion between collisions is free we can explicitly give the full 
orbit from this sequence.  For billiards the first return map, 
$q_{n+1} = f(q_n,p_n), p_{n+1} = g(q_n,p_n)$, be obtained explicitly.  
This ``surface of section'' construction then reduces the problem to a 
simpler dynamics, that of a map, on a space of reduced dimensionality.

This construction is identical in spirit to the above interpretation 
of the pre--integral equation.  Rather than consider the 
space of all functions defined on $\triangle$ as dynamical variables 
we have reduced the dimensionality of the unknowns to the space of 
functions on $\partial\triangle$.  All we need an integral equation on 
this space to play the role of a ``quantum first return map'' and whose 
unique solution is the actual boundary normal derivative.

This analogy is of course not direct because of the fundamental 
differences between classical and quantum mechanics.  In quantum 
mechanics there is no sensible concept of ``phase space''.  The state 
of the system, in the Schroedinger form of quantum mechanics, involves 
only the interior and boundary configuration coordinates.  Furthermore 
$q$ is the independent variable of the wave function rather than a 
dynamical variable $q(t)$.

It would appear that a ``quantum first return map'' can be obtained in 
a straightforward manner by taking the gradient of 
\eqref{eq:PreIntegralEquation} with respect to $\vec{x}_f$, 
evaluating the limit as $\vec{x}_f$ tends to the boundary and 
projecting out the normal derivative.  We shall refer to this process 
as the \emph{boundary limit}. 

\subsection{The boundary limit}
 
The boundary limit is in fact rather subtle.  Let us now take up the 
standard approach in which the free Green's function is chosen for 
$\GeneralGF$.

\begin{figure}[htb]
    \centering
    \psfrag{0}[][]{$0$}
    \psfrag{1}[][]{$1$}
    \psfrag{2}[][]{$2$}
    \psfrag{P0}[][]{\Large $\vec{P}_0$}
    \psfrag{P1}[][]{\Large $\vec{P}_1$}
    \psfrag{P2}[][]{\Large $\vec{P}_2$}
    \psfrag{l0}[][]{$l_0$}
    \psfrag{l1}[][]{$l_1$}
    \psfrag{l2}[][]{$l_2$}
    \psfrag{phi0}[][]{$\phi_0$}
    \psfrag{phi1}[][]{$\phi_1$}
    \psfrag{phi2}[][]{$\phi_2$}
    \inputFigure{height=0.3\textheight}{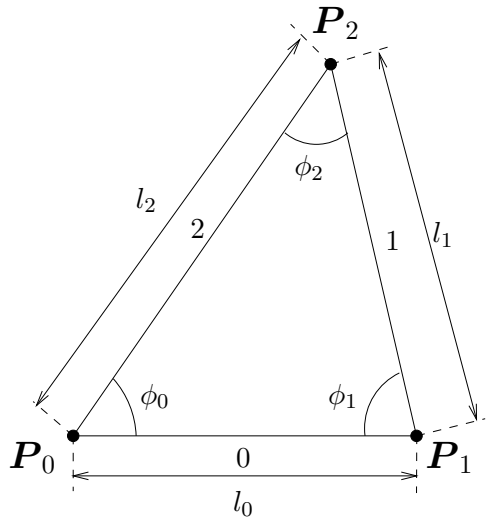}
    \caption{Notation: Free--based integral equation}
    \protect\label{fig:notationFree}
\end{figure}

At this point it is necessary to introduce some notation.  Refering to 
Figure~\ref{fig:notationFree} we locate the verticies of the 
triangle at $\vec{P}_0,\vec{P}_1$ and $\vec{P}_2$.  The sides 
of the triangle are labelled with the integers $0,1,2$ so that 
$\partial\triangle = 0 + 1 + 2$.  The length of side $\mu$ is 
$l_{\mu}$.

For convenience we parameterise each segment of the boundary using the 
distance from the vertex immediately clockwise:
\begin{subequations}
\begin{align}
  \vec{r}_{0}(s) 
  &= \vec{P}_0 + 
  \frac{\vec{P}_1-\vec{P}_{0}}{|\vec{P}_{1}-\vec{P}_{0}|}
  \,s \quad\quad 0 \leq s \leq l_0 \\
  \vec{r}_{1}(s) 
  &= \vec{P}_{1} + 
  \frac{\vec{P}_{2}-\vec{P}_{1}}{|\vec{P}_{2}-\vec{P}_{1}|}
  \,s \quad\quad 0 \leq s \leq l_1 \\
  \vec{r}_{2}(s) 
  &= \vec{P}_{2} + 
  \frac{\vec{P}_{0}-\vec{P}_{2}}{|\vec{P}_{0}-\vec{P}_{2}|}
  \,s \quad\quad 0 \leq s \leq l_2 
\end{align}
\end{subequations}
Then we define 
\begin{subequations}\label{eq:algebraicNotation}
\begin{align}
  &&&&
  X_{\mu}(t)          
  &=
  \nabla T(\vec{r}_{\mu}(t), \vec{x}_i)\DotProd\vec{n}_{\mu}
  && 0 < t < l_{\mu}
  \\
  &&&&
  X^{0}_{\mu}(t)      
  &=
  2\nabla G_{0}(\vec{r}_{\mu}(t), \vec{x}_i)\DotProd\vec{n}_{\mu} 
  && 0 < t < l_{\mu}
  \\
  &&&&
  K_{\mu \nu}(t,s) 
  &=
  - 2\nabla 
  G_{0}(\vec{r}_{\nu}(s),\vec{r}_{\mu}(t))
  \DotProd\vec{n}_{\mu}
  && 0 < t < l_{\mu}\,, \quad 0 < s < l_{\nu}
  \end{align}
\end{subequations}
where $\mu = 0,1,2$, $\nu = 0,1,2$. 
The need for the mysterious factors of 2 will soon be divined.

Formally taking the boundary limit towards side $0$, we obtain
\begin{equation}\label{eq:FormalBL}
\begin{split}
	2 X_{0}(t) & = X^{0}_{0}(t) 
	 + \int_{0}^{l_{0}} K_{00}(t,s)X_{0}(s)\,ds\\
	&\quad\quad + \int_{0}^{l_{1}} K_{01}(t,s)X_{1}(s)\,ds +
	    \int_{0}^{l_{2}} K_{02}(t,s)X_{2}(s)\,ds
\end{split}
\end{equation}

\begin{figure}[htb]
    \centering
    \psfrag{t}[][]{$t$}
    \psfrag{s}[][]{$s$}
    \psfrag{i}[][]{$\mu$}
    \psfrag{j}[][]{$\nu$}
    \psfrag{X}[][]{$X_{\mu}(t)$}
    \psfrag{X0}[][]{$X_{\mu}^0(t)$}
    \psfrag{K}[][]{$K_{\mu\nu}(t,s)$}
    \inputFigure{}{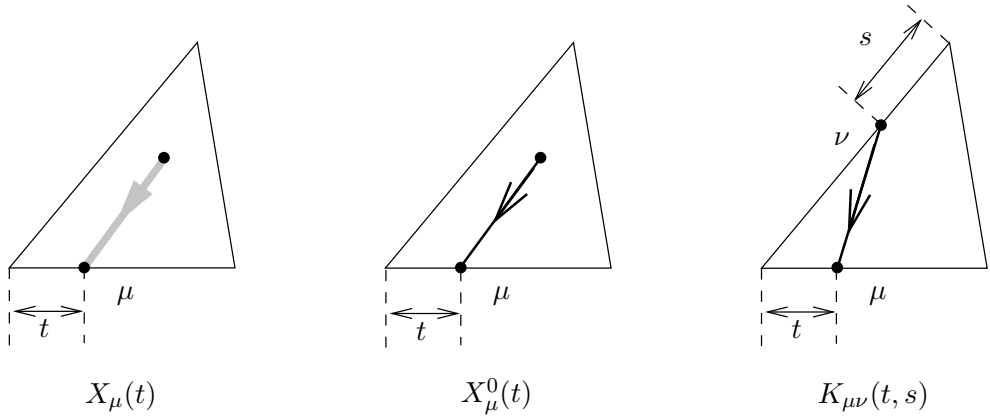}
    \caption{Feynman Rules for translating between integral equations and
    diagrammatic representations. The arrows give the ``flow'' from 
    $\mathbf{x}_i$ to $\mathbf{x}_f$ in the integral equations.
    They are needed to distinguish between $K_{\mu\nu}(t,s)$
    and $K_{\nu\mu}(t,s)$.
    }
    \protect\label{fig:FeynmanRules}
\end{figure}

In order to examine the boundary limit in detail we first present a 
diagramatic notation for \eqref{eq:FormalBL} using the ``Feynman 
rules'' in Figure \ref{fig:FeynmanRules}.  Diagrammatic notation is 
totally equivalent to the algebraic notation and provides a convenient 
tool for generating perturbation expansions.  In the present context 
it clearly elaborates the structure of the 
integral equations.  It will also prove very useful and natural in the 
generation of periodic orbit expansions.

The translation between notations is given by the usual rules
\begin{itemize}
	\item  Associate each factor in the integrand with an edge as given 
	in Figure \ref{fig:FeynmanRules}.
	
    \item  The configuration space dimensions in the arguments of the 
    integrand functions can be labelled on the triangle as shown. 
    However these are fairly obvious and usually omitted.
    	
	\item  An integration along a side is associated with each vertex
	where two edges meet.

\end{itemize}

\begin{figure}[htb]
    \centering
    \psfrag{t}[][]{$t$}
    \psfrag{s}[][]{$s$}
    \psfrag{2}[][]{$2$}
    \psfrag{e}[][]{$=$}
    \psfrag{p}[][]{$+$}
    \psfrag{IK}[l][l]{$\overbrace{\int_0^{d_1}ds}
                       \underbrace{K_{01}(t,s)}
                       \overbrace{X_1(s)}$}
    \inputFigure{}{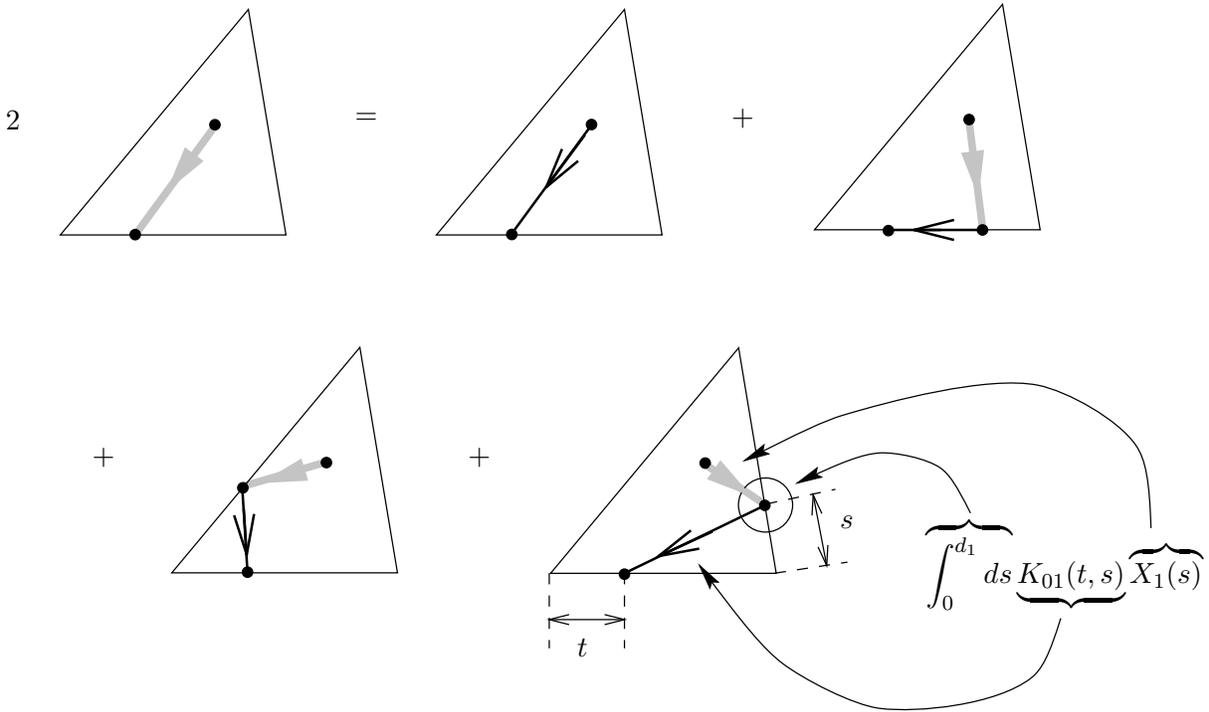}
    \caption{Graphical representation integral equation in naive boundary
    limit. Also shown is an example of the application of the Feynman rules.
    }
    \protect\label{fig:IntEqn}
\end{figure}

In Figure~\ref{fig:IntEqn} we give a graphical representation of 
\eqref{eq:FormalBL}.  The application of the Feynman rules is shown 
for part of the equation.

At this stage we require the explicit form of the integral kernel
\begin{equation}  \label{eq:StandardKernel}
  K_{\mu \nu}(t,s) 
  =
  -\frac{i\sqrt{z}}{2}
  H_{1}^{(1)}(\sqrt{z}|\vec{\Delta}_{\mu\nu}(t,s)|)
  \frac{\vec{\Delta}_{\mu\nu}(t,s)\cdot \vec{n}_{\mu}}
  {|\vec{\Delta}_{\mu\nu}(t,s)|} 
\end{equation}
Here
$\vec{\Delta}_{\mu\nu}(t,s) = \vec{r}_{\mu}(t) - \vec{r}_{\nu}(s)$ and 
$H_{1}^{(1)}(z)$ is a first order Hankel function of the first 
kind~\cite{Abramowitz68}.

The subtlety in the boundary limit arises because
\begin{equation}\label{eq:ElectrostaticLimit}
	K_{\mu \nu}(t,s) 
	\approx 
	- \frac{1}{\pi}
	\frac{\vec{\Delta}_{\mu\nu}(t,s)\DotProd\vec{n}_{\mu}}
    {|\vec{\Delta}_{\mu\nu}(t,s)|^{2}}  
	\quad |\vec{\Delta}_{\mu\nu}(t,s)| \ll 
	\frac{1}{\sqrt{z}}
\end{equation}

Apart from this the kernel is well behaved.  The singularity occurs 
when $|\vec{\Delta}_{\mu\nu}(t,s)|$ is much smaller 
than the de Broglie wavelength so that it appears in the 
\emph{quantum} regime. This distance is explicitly represented in the 
diagrammatic notation by the length of the segment corresponding to the 
``free propagator'', $K_{\mu\nu}$.  It is thus easy to see when a singularity 
is encountered in the integration.

Let us now go back and do the boundary limit towards side $0$ in a 
more careful manner.  In boundary limit we take $\vec{x}_{f} \to 
\vec{r}_{0}(t)$. Let us make the limit explicit by letting
$\vec{x}_{f} = \vec{r}_{0}(t) - \epsilon\,\vec{n}_{0}$.  We can readily 
``deform'' our diagrammatic notation to accomodate this.  In 
particular, in Figure \ref{fig:IntEqn} we can denote this deformation 
by simply displacing the final point slightly above the boundary.

It is easy to see that the boundary limits of the left hand side and 
the inhomogenous term exist by continuity, since $\vec{x}_{i}$ is 
strictly in the interior.  It is also clear that, provided we stay 
away from the corners to keep the distance of free propagation bounded 
from below, the integrations over boundaries $1$ and $2$ have uniformly 
bounded integrands.  They are then uniformly convergent and 
the boundary limit can be taken inside the integrals to give the 
formal result.  The same is not true for the integration over boundary 
$0$.  Here it is clear that the integrand develops a singularity in 
the boundary limit due to the presence of arbitrarily short distance 
propagation.  The consequent lack of uniform convergence means that 
the formal boundary limit is not correct.  This observation has 
already been made by Li and Robnik~\cite{LiRobnik95}.  Their remedy is 
to perturb about the half--plane.  Their treatment fits into our 
general scheme and we shall describe it in 
Section~\ref{sec:HalfPlane}.

The boundary limit towards side $0$ can still be done, but care is required. 
In explicit form, we need to calculate.
\begin{equation}
	\lim_{\epsilon\to0}
	\int_{0}^{l_{0}}
	\frac{i\sqrt{z}}{4}
	H_{1}^{(1)}(\sqrt{z}\sqrt{(t-s)^{2} + \epsilon^{2}})
	\frac{\epsilon}{\sqrt{(t-s)^{2} + \epsilon^{2}}}
	X_{0}(s)\,ds
\end{equation}
The integrand tends to zero as $\epsilon\to0$ for $t \neq s$.  Because 
the singularity in \eqref{eq:ElectrostaticLimit} reflects the unit 
source condition satisfied by the free Green's function we expect to 
see $\delta$--function behaviour.  This can be revealed explicitly by 
scaling the dummy variable according to $s = t + \epsilon x$ to obtain
\begin{equation}\label{eq:CarefulBL}
	\lim_{\epsilon\to0}
	\int_{-t/\epsilon}^{(l_{0} - t)/\epsilon}
	\frac{i\sqrt{z}}{4}
	H_{1}^{(1)}(\sqrt{z}\epsilon\sqrt{x^{2} + 1})
	\frac{1}{\sqrt{x^{2} + 1}}
	X_{0}(t + \epsilon x)\,\epsilon\,dx
\end{equation}
Expanding $X_{0}(t+\epsilon x)$ as a Taylor series in $\epsilon$ and 
using the small argument form for the Bessel function we can 
explicitly evaluate the integral for small $\epsilon$ to obtain
\begin{equation}
	\left[
	\lim_{\epsilon\to0}
		\frac{1}{2\pi}
		\int_{-t/\epsilon}^{(l_{0} - t)/\epsilon}
		\left\{ 
			\frac{1}{x^{2} + 1}
			+ O(\epsilon \log\epsilon)
		\right\}
		\,dx
	\right]
    X_{0}(t) 
	= \frac{1}{2}X_{0}(t)
\end{equation}
We see then that in the boundary limit the integrand behaves as 
$\frac{1}{2}\delta(x)$.

Repeating the calculation for the boundary limit toward sides $1$ and 
$2$ (or using a cyclic permutation of labels), we obtain

\begin{subequations}\label{SingularIntegralEquation}
\begin{align}
	X_{0}(t) &= X^{0}_{0}(t) 
		+ \int_{0}^{l_{1}} K_{01}(t,s)X_{1}(s)\,ds +
	    \int_{0}^{l_{2}} K_{02}(t,s)X_{2}(s)\,ds
	\\
	X_{1}(t) &= X^{0}_{1}(t) 
		+ \int_{0}^{l_{0}} K_{10}(t,s)X_{0}(s)\,ds +
	    \int_{0}^{l_{2}} K_{12}(t,s)X_{2}(s)\,ds
	\\
	X_{2}(t) &= X^{0}_{2}(t) 
		+ \int_{0}^{l_{0}} K_{20}(t,s)X_{0}(s)\,ds +
	    \int_{0}^{l_{1}} K_{21}(t,s)X_{1}(s)\,ds
\end{align}
\end{subequations}
This is a coupled form of the standard integral equation.

From \eqref{eq:StandardKernel}
we observe that $K_{00}(t,s) = 0$ because $\vec{r}_{0}(t) - 
\vec{r}_{0}(s) \perp \vec{n}_{0}$.  It would be quite easy, in a 
first treatment, to ignore the fact that this is strictly only true 
for $t \neq s$ and use it in the formal boundary limit.  This would 
lead to the same set of integral equations, except that the factors of 
2 in \eqref{eq:algebraicNotation} would be absent.  We made this 
mistake in our early investigations and are quite confident that in 
doing so we have tread a well worn path.

\subsection{Other treatments of the boundary limit}

Balian and Bloch~\cite{BalianBloch}[I] derived the standard integral 
equations using potential theory.  Their pre--integral equation 
(II.10) with (II.12) is of a different form to ours due to the choice 
of a double layer potential.  We cannot understand how such an object 
can be obtained from Green's theorem.  The multiple scattering series 
which they give is nevertheless identical to the one obtained by 
iterating our equations.  Potential theory is essentially the 
application of Green's theorem to electrostatics.  Because the Poisson 
equation is very different to the Helmholtz equation we feel that, 
while potential theory may give correct results, the delicacy of the 
problem demands the direct use of Green's theorem.
 
Other authors~\cite{SieberPhD,BerryWilkinson} derive the standard 
integral equations from the Helmholtz form of Greens' second identity.
\begin{equation}\label{eq:HelmholtzGreen}
	\int_{\partial D}ds'
		\left\{
		\psi(\vec{r}')\,
		\vec{n'}\DotProd\nabla_{\vec{r}'}
		G_{0}(\vec{r},\vec{r}') - 
		G_{0}(\vec{r},\vec{r}')\,
		\vec{n'}\DotProd\nabla_{\vec{r}'}
		\psi(\vec{r}')
		\right\}
		= \begin{cases}
				\psi(\vec{r}) & \vec{r} \in D/\partial D\\
				\frac{1}{2}\psi(\vec{r}) & \vec{r} \in \partial D\\
				0 & \vec{r} \not\in D
			\end{cases}
\end{equation}
The mysterious factor of $\frac{1}{2}$ can be understood as being due 
to the indentation of the contour with a semicircle when $\vec{r}$ 
appears on the boundary.

The derivation proceeds by differentating \eqref{eq:HelmholtzGreen}. As our 
discussion shows, there is clearly some sort of boundary layer 
behaviour so that \eqref{eq:HelmholtzGreen} needs to be 
supplemented with a statement of this behaviour. We suspect that performing 
the differentiation \emph{within} the boundary layer is justified. As 
such we feel that Li and Robnik~\cite{LiRobnik95} are not strictly 
correct in describing the approach as in error. As the reader now has 
our discussion of the boundary limit, as well as the approach or Li 
and Robnik, we feel no particular need to elaborate any further.

Our study the boundary limit was motivated by the hope, 
forlorn as it turned out, that by doing the problem ``carefully'' the 
singular nature of the integral equations could be removed.

\subsection{Singular nature: Fredholm theory}

In taking the boundary limit it is convenient to use a notation 
which identifies each side of the triangle explicitly.  This coupled 
form is convenient practical calculations because it provides
a \emph{symbolic dynamics} for \emph{quantum progagation}.  In the 
semiclassical analysis this will reduce to the natural symbolic 
dynamics for the classical propagation.

For theoretical purposes, in particular to apply the Fredholm theory, 
it is 
convenient to convert these coupled integral equations into a single 
integral equation.  To do this we first define the cumulative arc 
length to each vertex from $\vec{P}_{0}$: $\sigma_{0}=0,
\sigma_{\mu + 1} = \sigma_{\mu} + l_{\mu}$,  $\mu = 1,2,3$. 
We can then define
$\vec{x}(\sigma) = \vec{r}_{\mu}(\sigma - \sigma_{\mu})$ and 
$\vec{n}(\sigma) = \vec{n}_{\mu}$ for $\sigma_{\mu} < \sigma <
\sigma_{\mu+1}$ giving the position and normal vectors in the 
Birkhoff co--ordinate $\sigma$.

Then we define
\begin{subequations}
\begin{align}
	X(\sigma) & =  
		\nabla T(\vec{x}(\sigma), \vec{x}_{i})\DotProd\vec{n}(\sigma)
	\\
	X_{0}(\sigma) & =  
		2\nabla G_{0}(\vec{x}(\sigma), \vec{x}_{i})\DotProd\vec{n}(\sigma)
	\\
	K(\tau,\sigma) & =  
		-2\nabla G_{0}(\vec{x}(\tau),\vec{x}(\sigma))\DotProd\vec{n}(\tau)
\end{align}
\end{subequations}
and rewrite \eqref{SingularIntegralEquation} in the more compact form
\begin{equation}\label{eq:StandardCoupledForm}
	X_{\mu}(t) = X_{\mu}^{0}(t) + 
		\sum_{\nu = 0}^{2}
			\int_{0}^{l_{\mu}} K_{\mu\nu}(t,s) X_{\mu}(s)\,ds
\end{equation}
where we have redefined $K_{\mu\mu}(t,s) \equiv 0$. This plays the 
role of a \emph{quantum pruning rule}, resulting from the boundary 
limit process.
We then deduce
\begin{subequations}\label{eq:Translation}
\begin{align}
	K_{\mu\nu}(t,s) & =  K(\sigma_{\mu} + t, \sigma_{\nu} + s)
	\\
	X_{\mu}(t) & =  X(\sigma_{\mu} + t)
	\\
	X_{\mu}^{0}(t) & =  X^{0}(\sigma_{\mu} + t)
\end{align}
\end{subequations}
\eqref{eq:StandardCoupledForm} can then be translated into an uncoupled form
\begin{equation}
	X(\tau) = X^{0}(\tau) + \int_{\sigma_{0}}^{\sigma_{3}} K(\tau,\sigma) 
	X(\sigma)\,d\sigma
\end{equation}

In this form we can readily apply the Fredholm Theory. The important 
quantity here is the \LSquared norm of the kernel.
\begin{equation}
	\|K\| = \sqrt{\mathrm{Tr}(KK^{\dag})} 
		  = \sqrt{
			\int_{\sigma_{0}}^{\sigma_{3}}\,d\sigma
			\int_{\sigma_{0}}^{\sigma_{3}}\,d\tau
			|K(\tau,\sigma)|^{2}
			}
	\label{eq:NormDefn}
\end{equation}

where $^{\dag}$ denotes the Hermitian adjoint. 
In order to estimate the norm we revert to the coupled form using 
\eqref{eq:Translation}
\begin{equation}
	\sum_{\mu = 0}^{2}
	\sum_{\nu = 0}^{2}
		\int_{0}^{l_{\mu}}\,dt
		\int_{0}^{l_{\nu}}\,ds
			|K_{\mu\nu}(t,s)|^{2}
\end{equation}

The norm is the sum of 6 positive numbers.  It is clear from the 
diagrammatic notation that all 6 terms contain short distance 
propagation.  Let us now show that these short distance contributions 
cause the norm to diverge.  Consider the case
\begin{equation}
	\int_{0}^{l_{0}}\,ds
	\int_{0}^{l_{2}}\,dt
			|K_{20}(t,s)|^{2}
\end{equation}
The other 5 cases will follow similarly.  Let $t \to l_{2} - t$ so that both 
distances are refered to vertex $0$.

The integration region consists of the rectangle 
$(0,l_{0})\otimes(0,l_{2})$.  In this rectangle draw a quarter circle 
of radius $\delta \ll 1/\sqrt{z}$ centered on the origin.  This 
divides the integration into two domains.  In the bounded domain 
outside the quarter circle the integrand is bounded and continuous so 
that the integral exists.  Further subdivide the remaining quarter 
circle using a smaller quarter circle of radius $\epsilon$ centered on 
the origin. Consider the integration over the quarter annulus so 
defined.
Because $\Delta \ll 1/\sqrt{z}$ in this region we can use 
\eqref{eq:ElectrostaticLimit}.  Using polar 
coordinates in the $(t,s)$ plane one readily shows that the 
integration over the annular region is
\begin{equation}\label{eq:NormSingularity}
	\frac{1}{\pi^{2}}\,
	I(\phi_{0})\,
	\log \frac{\delta}{\epsilon} 
\end{equation} 
where $I(\phi_{0})$ is some unimportant integral.

The norm diverges logarithmically and \emph{the Fredholm theory does 
not apply}.  In fact a portent of the singular nature appears in a 
more careful examination of the boundary limit.  In 
\eqref{eq:CarefulBL} it was implicitly assumed that $l_{0} - t$ and 
$t$ were not $O(\epsilon)$.  If we had taken the boundary limit whilst 
keeping $t/\epsilon$ constant we would have obtained a different, 
\emph{$\epsilon$--dependent} result. (This gives a quantitative measure of 
what we meant by the phrase ``provided we stay away from the 
corners''.) This involves the wave function a distance 
$\epsilon$ \emph{into the interior} of the domain so that 
resulting integral equation is defined on $\triangle$ rather than 
$\partial\triangle$ and is not the ``quantum first return 
map'' we seek.  The reader may be able to convince herself that there 
is no direct remedy for this boundary layer behaviour.

\subsection{Perturbative character of boundary integral method}

Confronted with no apparent way to overcome this singular behaviour 
in a mathematically satisfactory manner, one is forced to reassess the 
derivation. The first thing the reader must appreciate is that the 
method presented is a \emph{perturbative} method. 

In quantum and classical mechanics complicated Hamiltonians are 
treated by spliting off an ``interaction'' term $V$ so that $H = H_{0} 
+ V$ and the problem corresponding to $H_{0}$ can be explicitly 
solved.  One then constructs a perturbation series based on this 
solution.  The conversion of the boundary value problem to an integral 
equation presented above is directly analogous to this common 
technique.  Here the simplification is not achieved by spliting the 
Helmholtz operator but by simplifying the boundary conditions.  In an 
operator theoretic sense there is no distinction between the two 
simplifications.  The result, however, is not a perturbation 
series but an integral equation so that it is a 
\emph{non--perturbative} result from a method with perturbative 
character.

The standard derivation obtains an ``$H_{0}$'' with 
a very brutal reduction of the boundary value problem.
Our investigations have led us to conjecture that:
\begin{quote}
 \emph{It is in principle not possible to obtain the full triangle Green's 
function perturbing about the free Green's function}
\end{quote}
If this conjecture is correct then the next step is obvious: 
\begin{quote}
\emph{One 
must perturb about a Green's function which reflects more closely the 
basic structure of the full solution}
\end{quote}
At this point we require a catalog of exactly known Green's functions.
\begin{itemize}

\item For domains which tile Euclidean space under reflection, the 
      Greens function can be constructed by the method of images.

\item For the wedge Sommerfeld generalised the method of images to
      give the Greens function as a contour integral.

\item For the circle and other separable problems one can obtain 
      eigenfunction series for the Greens function. Semiclassical 
      studies then require use of complex angular momentum techniques
      (aka. Sommerfeld--Watson transform, Regge poles\cite{Wirzba92}). 

\end{itemize}

The reader familiar with these solutions is in a position to 
appreciate that the availability and relative simplicity of the 
method of images solutions and the wedge solution was the prime 
motivation in our decision to study the quantum triangular billiard.

\subsection{Perturbing about the half--plane}
\label{sec:HalfPlane}

The next simplest Green's function about which to perturb is the 
Dirichlet Green's function on the half--plane $D_{H}$
%
%
\begin{equation}
	H(\vec{x}, \vec{x}_f) = 
	G_0(\vec{x}, \vec{x}_f) - 
	G_0(\vec{x}, \vec{x}_f^R),  
\end{equation}
%
%
obtained using the method of images. $\vec{x}_f^R$ is 
the mirror image of $\vec{x}_f$ in the boundary $\partial 
D_{H}$.

Li and Robnik~\cite{LiRobnik95} have presented a half--plane based 
derivation for smooth billiards and shown that the standard 
integral equations are recovered.  Let us briefly present this 
derivation for the triangle both as a warmup to the wedge case as 
well as an elaboration of Li and Robnik's result.

First extend side $0$ to infinity in both directions.  Denote the 
half--plane Green's function which vanishes on this line by 
$H_0(\vec{x},\vec{x}_f)$
\footnote{As this is the only place where the half--plane solution is 
used, no confusion with the notation for the Hankel functions should 
arise}. 
Using this Green's function in \eqref{eq:PreIntegralEquation} gives
\begin{equation}    \label{eq:HalfPlanePreIntegralEquation}
  T(\vec{x}_f,\vec{x}_i) = 
    H_0(\vec{x}_i,\vec{x}_f) -
    \int_{1 + 2}
       H_0(\vec{x},\vec{x}_f) 
       \nabla \FullGF \DotDNormal   
\end{equation}
The Dirichlet boundary condition satisfied by $H_0(\vec{x},\vec{x}_f)$ 
eliminates the delicate integration over side $0$.  Because of this 
the boundary limit becomes straightforward, provided, once again, that 
we are not too close to the corners. Perturbing about the 
half--plane Green's functions which vanish on sides $1$ and $2$ and 
taking the boundary limit we recover the standard integral equations
\eqref{SingularIntegralEquation}. The ``mysterious'' factors of 2 in 
\eqref{eq:algebraicNotation} arise here because the image 
term gives a contribution to the kernel which is identical to that of 
the free term. 

Note that Sieber uses a perturbation about the quadrant Green's 
function in an application of the boundary integral method to the 
hyperbola billiard\cite{SieberPhD}.

\subsection{Direct removal of singular nature}
\label{sec:DirectFaddeev}

It is possible to derive a non--singular integral equation from the 
singular integral equation.  Riddell~\cite{Riddell} does this by 
separating out the singular part of the kernel and treating it exactly 
via an auxiliary integral equation.  Riddells' technique is somewhat 
cumbersome and is not particularly intuitive.  
Faddeev~\cite{Joachain79}, in his treatment of the quantum three body 
problem, used a more systematic version of this approach.

We began our studies trying to following Faddeev's method. We were able to
successfully adapt Faddeev's techniques to our present situation by
splitting the (matrix) kernel into three pieces where each piece involved
only collisons between two sides. This is similar to the decomposition of
the potential energy in the three body problem as
$V_{12} + V_{23} + V_{31}$. 
One can then follow Faddeev fairly closely, defining T-matricies
for each ``wedge'' of the decomposition and using these to derive
non--singular integral equations. We found the result
unsatisfactory because we were unable to obtain a convenient form for the
T-matricies.

\section{Wedge--Based Approach: Non-Singular Integral Equations}
\label{sec:WedgeBased}

The next most natural way to proceed is to base a derivation on the 
Dirichlet Green's function of the wedge.

\subsection{Historical overview}

The scattering problem for the wedge is of considerable importance in 
optics where it is virtually the only non--trivial example in which 
Maxwell's equations can be solved exactly~\cite{BornWolf80}.  It was 
first considered by Sommerfeld who approached it via a generalisation 
of the method of images to Riemann surfaces.  Sommerfeld's techniques 
were simplified and applied to other problems by 
Carslaw~\cite{Carslaw99}.  It is a classical problem in mathematical 
physics and is treated in many text books as well as in the research 
literature~\cite{WedgeReview86}.  Attention is often given to the 
special case of the edge where the wedge angle is $2\pi$.

The wedge solution has been used as the basis for analysis in a number 
of studies.  Kac~\cite{KacDrum66} used the solution of the diffusion 
equation on a wedge to determine the constant term in Weyl's formula 
for the density of states.  More recently Aurell and 
Salmonson~\cite{AurellSalomonson94} have presented a similar argument 
in their study of the determinant of the 
Laplacian on triangular domains.  
The wedge solution has also appeared in the recent work of Pavloff and 
Schmit~\cite{PavloffSchmit95} which incorporates wedge diffraction in 
periodic orbit expansions for the triangle.  Diffraction effects in 
periodic orbit expansions were first considered by Vattay, Wirzba and 
Rosenqvist~\cite{VattayPRL94} using Keller's geometrical theory of 
diffraction~\cite{KellerGTD62}.  In these studies scattering off 
verticies involves a factor which originates in the asymptotic form of 
the scattering solution for the wedge .

In our case we require the Greens function for the wedge satisfying a 
unit source condition rather the condition of an ``outgoing wave'' at 
infinity.  This can be obtained using Sommerfeld's 
method~\cite{Carslaw18} although it was originally obtained by 
Macdonald in 1902 using other methods~\cite{Macdonald15}.  The 
solution is of same form as that of the scattering problem except that 
the plane wave is replaced by a Bessel function of a somewhat more 
complicated argument.  Because the Green's function has been considered 
to a lesser degree we shall give some details.  Despite appearances it 
is just as workable as the scattering solution.  In the 
appendix shall give an overview of its derivation.

\subsection{Naive, wedge--based integral equations.}

Refering to Figure \ref{fig:notationFree} let us label the wedge solutions 
according the vertex they share with the triangle.  Perturbing about 
$W_{0}(\vec{x}, \vec{x}_{f})$ and using \eqref{eq:PreIntegralEquation}
gives the pre--integral equation
\begin{equation} 
	\FullGF =
	W_{0}(\vec{x}_{i},\vec{x}_{f}) -
	\int_{1}
	W_{0}(\vec{x},\vec{x}_{f})\nabla\FullGF\DotDNormal   
\end{equation}
where the Dirichlet boundary condition satisfied by the wedge removes 
\emph{two} integrations.

\begin{figure}[htb]
    \centering
    \inputFigure{height=0.3\textheight}{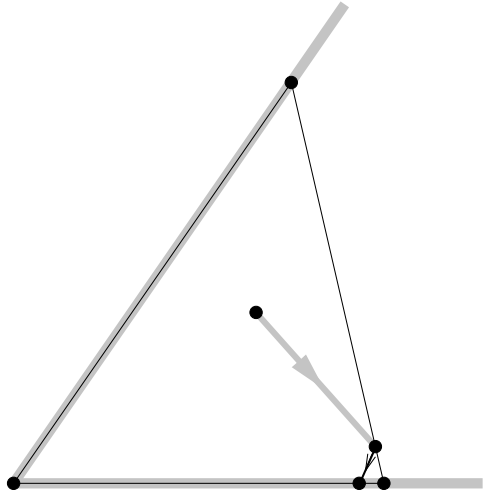}
    \caption{Corner divergences in the naive wedge--based derivation:
perturbation about the wedge solution based on vertex 0.}
    \protect\label{fig:WedgeCornerProblems}
\end{figure}

One can now take the boundary limit to either side 0 or side 2 without 
encountering the short distance singularity in the propagator.  
However as Figure~\ref{fig:WedgeCornerProblems} illustrates, taking 
the limit too close to corners 1 or 2 produces short distance 
propagation.  Because the wedge solution satisfies a unit source 
condition its short distance behaviour is the same as the free Greens 
function so that, once again, short distance propagation causes the 
norm of the resulting kernel to diverge.

Carrying through the derivation regardless, one obtains 6 
coupled equations for 3 unknown functions. This is an overspecified 
system, as is most readily seen by discretising the result to 
obtain a problem of linear algebra. It is possible to remove the 
short distance problems by taking advantage of this overspecification.

\subsection{``Regularising'' the singular nature} 

\begin{figure}[htb]
    \centering
    \psfrag{0}[][]{$0$}
    \psfrag{1}[][]{$1$}
    \psfrag{2}[][]{$2$}
    \psfrag{3}[][]{$3$}
    \psfrag{4}[][]{$4$}
    \psfrag{5}[][]{$5$}
    \psfrag{P0}[][]{\Large $\vec{P}_0$}
    \psfrag{P1}[][]{\Large $\vec{P}_1$}
    \psfrag{P2}[][]{\Large $\vec{P}_2$}
    \psfrag{P3}[][]{\Large $\vec{P}_3$}
    \psfrag{P4}[][]{\Large $\vec{P}_4$}
    \psfrag{P5}[][]{\Large $\vec{P}_5$}
    \psfrag{l0}[][]{$l_0$}
    \psfrag{l1}[][]{$l_1$}
    \psfrag{l2}[][]{$l_2$}
    \psfrag{l3}[][]{$l_3$}
    \psfrag{l4}[][]{$l_4$}
    \psfrag{l5}[][]{$l_5$}
    \psfrag{phi0}[][]{$\phi_0$}
    \psfrag{phi2}[][]{$\phi_2$}
    \psfrag{phi4}[][]{$\phi_4$}
    \inputFigure{height=0.3\textheight}{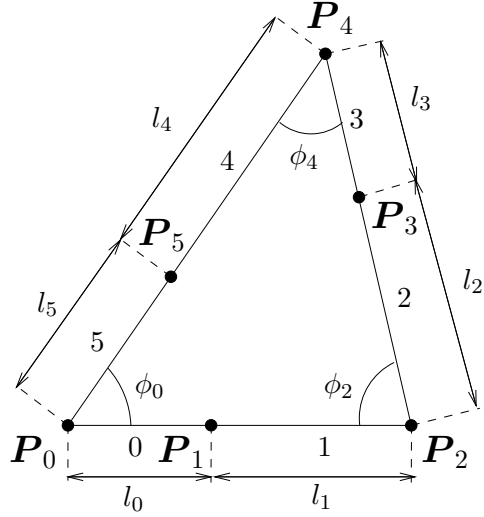}
    \caption{Notation: Wedge--based integral equation}
    \protect\label{fig:notationWedge}
\end{figure}

We begin by adding an extra vertex on each side of the triangle so 
that Figure~\ref{fig:notationFree} becomes Figure~\ref{fig:notationWedge}. 
In this manner we regard the triangle as a degenerate 6--sided 
polygon. The manner in which the sides are subdivided is arbitrary 
but, as we will discuss later, there will be an optimal subdivision.

The boundary is now parameterised in 6 segments 
\begin{equation} \label{eq:BoundaryPara}
	\vec{r}_{\mu}(s) = \vec{P}_{\mu} + 
		\frac{\vec{P}_{\mu + 1} -\vec{P}_{\mu}}
			 {|\vec{P}_{\mu + 1} -\vec{P}_{\mu}|}\,s
	\quad 0 < s < l_{\mu} 
\end{equation}
where $\mu = 0,1,2,3,4,5$ and $\vec{P}_{6} \equiv \vec{P}_{0}$ in a natural 
cyclic manner.  We denote the outward normals by 
$\vec{n}_{0}=\vec{n}_{1}$, $\vec{n}_{2}=\vec{n}_{3}$ and 
$\vec{n}_{4}=\vec{n}_{4}$.

The three wedge solutions about which one can perturb are now 
associated with verticies $0$, $2$ and $4$.  Let us consider the 
perturbation about $W_{0}(\vec{x}, \vec{x}_{f})$. 

Define
\begin{subequations}\label{eq:algebraicNotationWedge}
\begin{align}
  &&&&
  X_{\mu}(t)          
  &=
  \nabla T(\vec{r}_{\mu}(t), \vec{x}_i)\DotProd\vec{n}_{\mu}
  && 0 < t < l_{\mu}
  \\
  &&&&
  X^{0}_{\mu}(t)      
  &=
  2\nabla W_{0}(\vec{r}_{\mu}(t), \vec{x}_i)\DotProd\vec{n}_{\mu} 
  && 0 < t < l_{\mu}
\end{align}
\end{subequations}
where $\mu = 0,1,2,3,4,5$ and
\begin{equation}
  K_{\mu \nu}(t,s) 
  =
  - 2\nabla 
  W_{0}(\vec{r}_{\nu}(s),\vec{r}_{\mu}(t))
  \DotProd\vec{n}_{\mu}
  \quad\quad\quad 0 < t < l_{\mu}\,, \quad 0 < s < l_{\nu}
\end{equation}
where $\mu = 0,1,4,5$ and $\nu = 2,3$.
\eqref{eq:PreIntegralEquation} then gives 
the pre--integral equation.
\begin{equation}    \label{eq:WedgePreIntegralEquation}
	T(\vec{x}_{f},\vec{x}_{i}) =
	W_{0}(\vec{x}_{i},\vec{x}_{f}) +
	\int_{0}^{l_{2}}
	W_{0}(\vec{r}_{2}(s),\vec{x}_{f})\,X_{2}(s)\,ds
	+
	\int_{0}^{l_{3}}
	W_{0}(\vec{r}_{3}(s),\vec{x}_{f})\,X_{3}(s)\,ds
\end{equation}

\begin{figure}[htb]
    \subfigure[Non--Singular boundary limit]{
    \begin{minipage}{0.45\textwidth}
    \centering
    \psfrag{0}[][]{$0$}
    \psfrag{1}[][]{$1$}
    \psfrag{2}[][]{$2$}
    \psfrag{3}[][]{$3$}
    \psfrag{4}[][]{$4$}
    \psfrag{5}[][]{$5$}
    \psfrag{P0}[][]{\Large $\vec{P}_0$}
    \psfrag{P1}[][]{\Large $\vec{P}_1$}
    \psfrag{P2}[][]{\Large $\vec{P}_2$}
    \psfrag{P3}[][]{\Large $\vec{P}_3$}
    \psfrag{P4}[][]{\Large $\vec{P}_4$}
    \psfrag{P5}[][]{\Large $\vec{P}_5$}
    \inputFigure{height=0.3\textheight}{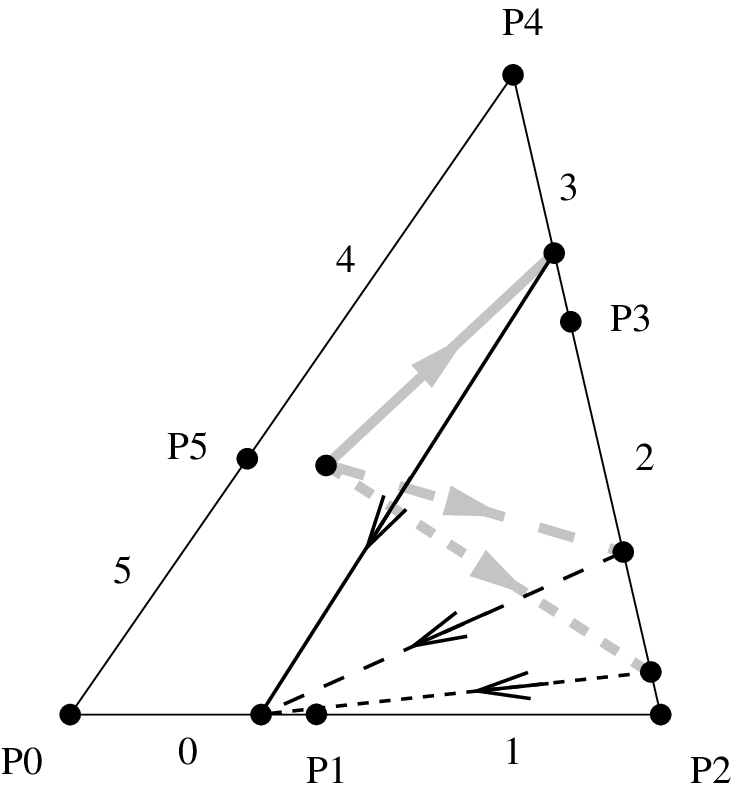}
    \end{minipage}
    }
    \subfigure[Singular boundary limit]{
    \begin{minipage}{0.45\textwidth}
    \centering
    \psfrag{0}[][]{$0$}
    \psfrag{1}[][]{$1$}
    \psfrag{2}[][]{$2$}
    \psfrag{3}[][]{$3$}
    \psfrag{4}[][]{$4$}
    \psfrag{5}[][]{$5$}
    \psfrag{P0}[][]{\Large $\vec{P}_0$}
    \psfrag{P1}[][]{\Large $\vec{P}_1$}
    \psfrag{P2}[][]{\Large $\vec{P}_2$}
    \psfrag{P3}[][]{\Large $\vec{P}_3$}
    \psfrag{P4}[][]{\Large $\vec{P}_4$}
    \psfrag{P5}[][]{\Large $\vec{P}_5$}
    \inputFigure{height=0.3\textheight}{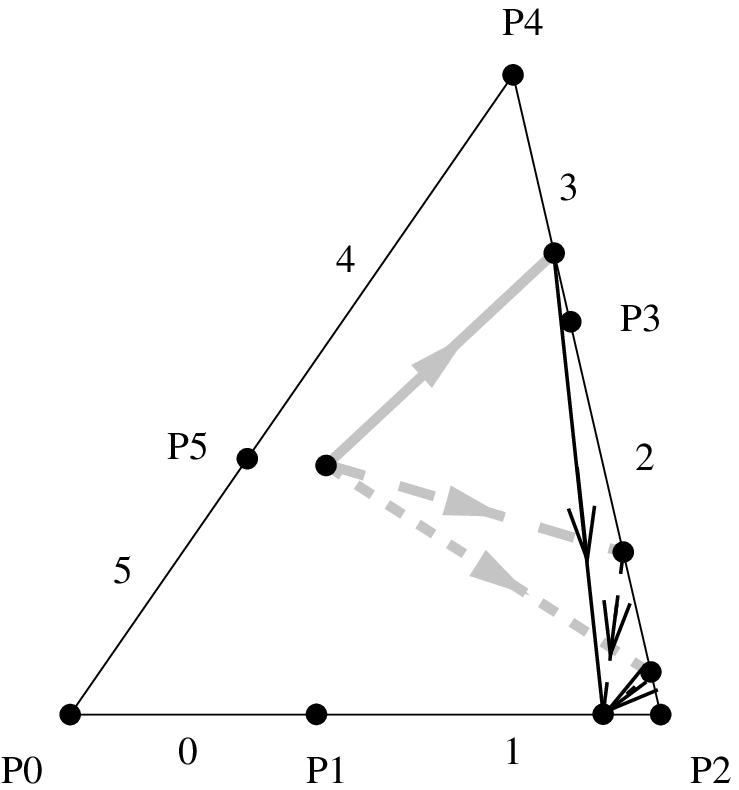}
    \end{minipage}
    }
    \caption{Boundary limit of wedge pre--integral 
      equation~\protect\eqref{eq:WedgePreIntegralEquation} to two typical
      segments. The integration over sides 2 and 3 is schematically given
      to show the absence (presence) of short distance propagation in 
      the boundary limit to 0 (1).
      }
    \protect\label{fig:WedgeBoundaryLimit}
\end{figure}

There are now 4 possible boundary limits.  Defining a diagramatic 
notation analogous to that given in Figure~\ref{fig:FeynmanRules} we 
illustrate 2 typical boundary limits in 
Figure~\ref{fig:WedgeBoundaryLimit}.  Taking the limit to sides $0$ 
and $5$ it is clear that there is a \emph{lower bound on the 
propagation distance}.  This is not the case taking the limit to sides 
$1$ and $4$. We take advantage of the overspecification, only using
the non--singular limits.

Perturbing about the wedge solutions associated with vertices $2$ and 
$4$ in an analogous manner gives a set of 
six coupled integral equations
\begin{equation} \label{eq:NonSingularIE}
	X_{\mu}(t) = X_{\mu}^{0}(t) 
		+ \sum_{\nu = 0}^{5}\int_{0}^{l_{\nu}} 
		  K_{\mu\nu}(t,s)X_{\mu}(s)\,ds
\end{equation}
Of the 36 possible $K_{\mu\nu}$ only the following 12 
are non zero
\begin{subequations}
\begin{align}
         K_{\mu,\mu+2}, K_{\mu,\mu+3} & \quad\quad\mu \text{ even}
 \\
         K_{\mu,\mu+3}, K_{\mu,\mu+4}   & \quad\quad\mu \mbox{ odd }
\end{align}
\end{subequations}
where addition is modulo $6$.  These quantum pruning rules are 
easier to remember using the diagrammatic form of the integral 
equations.  The kernel $K_{\mu\nu}$ is constructed using the wedge 
solution whose vertex is closest to the side corresponding to the 
first index.

Perturbing about the wedge thus gives us a closed set of non--singular 
integral equations for the unknown normal derivative.  The Fredholm 
theory guarantees that these equations define the unknown normal 
derivative uniquely. Substituting this solution into the pre--integral 
equation then gives the solution of the boundary value problem
\eqref{eq:TriangleBVP}.

In a sense, the need to start from the wedge solution is quite 
appealing because it gives a ``bootstrap'' character to the solution 
of boundary value problems: the triangle is solved in terms of the 
wedge problem; the wedge in terms of the free Green's function.  One 
is led to think how the triangle solution might be used in 
constructing even more complicated solutions (perhaps the 
tetrahedron).

\subsection{Validation}

Although the derivation has been straightforward and the result is 
more or less clear, it is nevertheless important to validate our 
approach.  The first step towards this is to carry out a standard 
numerical discretisation of the integral equations, calculating the 
Fredholm determinant and thus the spectrum.  It is important to 
reproduce exactly known spectra and compare the results for general 
triangles with the singular approach.  We shall do this in 
section~\ref{sec:Numerics}.

In our numerical work we would like to be assured that any 
discrepencies, any problems, are entirely due to the formulation of 
the integral equations.  It is then important to be able to calculate 
the kernel of the wedge--based approach with well--controlled 
approximations.  This requires a detailed understanding of the wedge 
kernel.  It is to this that we now turn our attention.

\section{Wedge Kernel}
\label{sec:Kernel}

\begin{figure}[htb]
    \centering
    \psfrag{E1}[t][]{\footnotesize $-2\pi + i\alpha$}
    \psfrag{E2}[b][t]{\footnotesize $-2\pi - i\alpha$}
    \psfrag{E3}[t][]{\footnotesize $i\alpha$}
    \psfrag{E4}[b][t]{\footnotesize $-i\alpha$}
    \psfrag{E5}[t][]{\footnotesize $2\pi + i\alpha$}
    \psfrag{E6}[b][t]{\footnotesize $2\pi - i\alpha$}
    \inputFigure{height=0.3\textheight}{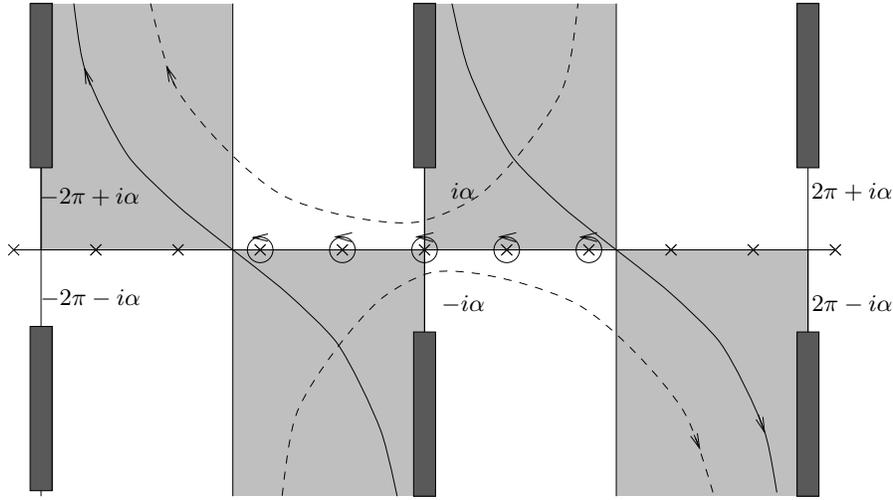}
    \caption{Wedge kernel: Analytic structure of integrand and integration
        contours. The integrand decays at infinity in the lightly shaded regions. 
        Heavy shading denotes branch cuts in $\lambda(\omega)$ and 
        crosses represent the poles of $P(\theta,\phi,\omega)$. 
        The dotted contour is the Carslaw contour $C$. For the
        semiclassical evaluation it is deformed to
        the solid contour which goes along the 
        paths of steepest descent though $\pm\pi$.
    }
    \protect\label{fig:WedgeContourOmega}
\end{figure}

In appendix~\ref{sec:wedgapp} we outline a derivation of the wedge 
Green's function and obtain the following form of the wedge 
kernel~\eqref{WedgeEq:Kernel}
\begin{equation} \label{eq:wedgeKernel}
\begin{split}
  K(r_{f},0; r_{i},\theta_{i};\phi)
  & \equiv 
  -\nabla G_{W}(r_{f},0; r_{i},\theta_{i})
	\cdot
	\left(-\hat{\theta}
	\right)
  \\
  & = \frac{1}{\phi r_{f}}\int_{C}^{} 
  P(\theta_{i},\phi;\omega)
  \frac{i}{4} H_{1}^{(1)}(\mu \lambda(\omega)) \mu \lambda'(\omega)
  d\,\omega
\end{split}
\end{equation}
where
\begin{subequations}
\begin{align}
  P(\theta_{i},\phi;\omega) & =
  \frac{1}{1 - e^{-i\pi (\theta_{i} + \omega)/\phi}} 
  \\
  \cosh\alpha & = 
  \frac{1}{2}
  \left( \frac{r_{f}}{r_{i}} + \frac{r_{i}}{r_{f}}
  \right) 
  \\	
  \lambda(\omega) & =  
  \sqrt{ \cosh\alpha - \cos \omega}
  \label{eq:LambdaDefinition}
  \\
  \mu & =  \sqrt{2z r_{i}r_{f}}
\end{align}
\end{subequations}
where the natural polar co--ordinate system has been used.  
$H_{1}^{(1)}$ is the first order Hankel function of the 1st kind.  The 
square root in \eqref{eq:LambdaDefinition} is the principal square 
root with a branch cut on the negative real axis.  The contour $C$ 
together with the analytic structure of the integrand is given in 
Figure~\ref{fig:WedgeContourOmega}.

Using
\eqref{eq:wedgeKernel} we have e.g., 
$K_{01}(t,s) =K(t,r_{i}(s),\theta_{i}(s);\phi_{0})$ 
where $r_{i}$ and $\theta_{i}$ 
are obtained using elementary trigonometry. We shall use the term 
``wedge kernel'' to refer to both \eqref{eq:wedgeKernel} and 
$K_{\mu\nu}(t,s)$ as there is little danger of confusion.

\subsection{How to calculate the contour integral}

As with all special functions, the calculation of \eqref{eq:wedgeKernel} has
to proceed through controlled approximations. In general, a given
approximation technique is only valid in a particular region of parameter
space so that one must consider a range of techniques in order to be able to
calculate the function globally. For \eqref{eq:wedgeKernel} we basically
have two regimes: the semiclassical regime,
$|\sqrt{z}(\mathbf{x}_{f} -\mathbf{x}_{i})| \gg 1$,
in which the method of steepest descents is suitable; 
and the quantum regime,  
$|\sqrt{z}(\mathbf{x}_{f} -\mathbf{x}_{i})| \ll 1$, 
where the standard eigenfunction series is suitable. By developing each
technique sufficiently one can extend each of them, with sufficient
computational effort, so that the two regimes overlap. We are then able to
calculate \eqref{eq:wedgeKernel} for all parameters. 

Naturally, we are initially interested in an effective semiclassical
approximation. Previous analyses \cite{Pauli38,Oberhettinger58} were
concerned with the scattering problem. The analysis for the
Greens function is similar. However, because we require a controlled
approximation it is neccessary to undertake a somewhat deeper treatment. 
Let us begin by outlining the standard semiclassical analysis.

\subsection{Asymptotic analysis and geometrical optics interpretation}

The semiclassical limit corresponds to $\mu\to\infty$.  In this limit 
the integrand is dominated by $e^{i\mu\lambda(\omega)}$ so that the 
method of steepest descents is appropriate.  One readily locates 
saddle points at $\pm\pi$ together with the local path of steepest 
descent.  Deforming the Carslaw contour 
$C$ to the steepest descent contours one collects contributions from 
the real poles in the region $-\pi < \Re(\omega) <\pi$ 
(Figure~\ref{fig:WedgeContourOmega}).  The pole contributions are of 
geometrical optics form, i.e.  a free Green's function in which the 
distance, $|\mathbf{x}_{f} - \mathbf{x}_{i}|$, is generalised to 
include paths from $\mathbf{x}_{i}$ to $\mathbf{x}_{f}$ via 
classically possible collisions with the walls.  All possible 
classical paths are represented with an appropriate phase for the 
corresponding free Green's function.  The remaining contour integrals, 
in the steepest descent approximation, can be interpreted in terms of 
a classical path from $\mathbf{x}_{i}$ to $\mathbf{x}_{f}$ via the 
vertex.  The importance of such paths, which are not classical but 
nevertheless minimise the action, was recognised by 
Keller~\cite{KellerGTD62} in his geometrical theory of diffraction.

The steepest descent integrations are not uniformly of Gaussian form.
The poles, located at $2n\phi - \theta_{i}$, $(n \in \field{Z})$, pass 
through the the steepest descent contour at certain values of 
$\theta_{i}$ so that the number of geometrical optics terms varies.  This 
simply corresponds to the motion of the ``shadow'' of the source 
point.  This apparently discontinuous behaviour is of course an 
artifact of the method of steepest descent.  It is well known that the 
geometric terms ``switch on'' with an error function modulation.  This 
is the Stokes' phenomenon and a uniform approach to it has been studied 
in recent years~\cite{BerryStokes91}.  A uniform approach to a special 
case of the scattering problem was given by Pauli~\cite{Pauli38}.

It would be natural at this point to derive asymptotic series.  This
only requires a \emph{local} knowledge of the integrand 
(in particular the steepest descent contour) at the 
saddle points and at infinity.  In order to determine the nature of 
the error and, if possible, bound it, this understanding needs to 
be \emph{global}. Because such an analysis is 
rarely presented we shall give it in some detail.

While we could in principle carry out an error analysis for the 
asymptotic series, this is not really neccessary for our 
purposes.  Asymptotic series are useful to understand the asymptotic 
form of the results.  We already understand this through the standard 
interpretation in terms of (generalised) classical paths.  
All we require is well--controlled approximations in order to 
validate the non--singular integral equation.  This is 
conveniently and practically obtained via Gauss-Hermite quadrature 
along the steepest descent contour.  Numerical quadrature along any 
other contour gives poor convergence due to oscillations in the 
integrand. (This is of course the rationale behind the steepest 
descent method.) 
We only know of one other study which uses the method of steepest descents 
as the basis of a numerical quadrature~\cite[p.265]{Carrier}.

\subsection{Global understanding of integrand: $\omega \leftrightarrow 
t$ transformation}  

The steepest descent contour $\omega(t)$ is 
defined implicitly by
\begin{equation}\label{eq:ContourDefinition}
	\lambda(\omega) = \lambda_{0} + \frac{it^{2}}{\mu},
	\quad\quad -\infty < t < \infty
\end{equation}
where $\lambda_{0} = \sqrt{\cosh\alpha + 1}$.  Rather than regard 
the variable $t$ as a \emph{parameterisation} of the steepest descent 
contour, 
consider~\eqref{eq:ContourDefinition} to define a \emph{transformation}
to a complex $t$--plane.  In the $t$--plane the contour is simply along 
the real axis. The integrand is of a semiclassical form only if it is 
a well peaked Gaussian in the $t$ variable.

Let us first write the conformal transformation 
\eqref{eq:ContourDefinition} in the following dimensionless manner
\begin{equation}
	\frac{i t^{2}}{\eta}
	= 
	\sqrt{1 - \frac{1 + \cos\omega}{\sigma}} - 1
\end{equation}
where
\begin{subequations}
\begin{align}
	\eta  & = \mu\lambda_{0} = \sqrt{z}(r_{i} + r_{f})
	\\
	\sigma & = \lambda_{0}^{2}  = 1 + \cosh\alpha
\end{align}
\end{subequations}
$\eta$ is essentially a dimensionless wave vector. The semiclassical limit is 
conventionally regarded as $\eta \gg 1$, corresponding to a de Broglie 
wavelength which is small compared to typical distances in the 
problem.  The parameter $\sigma$ is a purely geometrical, scale 
invariant parameter.  $\sigma \gg 1$ indicates that that one of the 
distances $r_{i},r_{f}$ is much larger than the other, i.e.  that one 
of them is ``close'' to the corner.  As we shall see the semiclassical 
limit requires not only the conventional $\eta \gg 1$ but also requires 
$\sigma \ll 1$.

$\omega \leftrightarrow t$ can be constructed via the sequence of 
transformations 
$\omega \leftrightarrow u$, 
$u \leftrightarrow v$, and
$v \leftrightarrow t$ where
\begin{subequations}
\begin{align}
	u & =  \frac{1 + \cos\omega}{\sigma}
	\\
	\frac{iv}{\eta} &= \sqrt{1 - u} - 1 
	\\
	t^{2} & = v
\end{align}
\end{subequations}
The choice of variable $u$ is suggested by the fact that $u=0$ at 
the saddle point and turns out to be convenient.  The other choices 
are more or less natural.

\begin{figure}[p]
  \subfigure[$\omega$--plane: Some features in $\omega\leftrightarrow u$
  are also shown (see Figure~\ref{fig:u}). The $n=0$ ($n=1$) branch of this
  transformation corresponds to $0 < \Re(\omega) < \pi$
  ($ \pi < \Re(\omega) < 2\pi$).
  The bold dotted lines are images of
  the imaginary $u$ axis. The bold solid lines are 
  images of the real $u$ axis. These lines partition the $\omega$--plane
  into regions. The correspondance between these regions and the quadrants
  of the $u$--plane is given by boxed numbers.
  ]{\begin{minipage}[b]{0.95\textwidth}
        \centering
        \psfrag{E1}[lt][lt]{\footnotesize $i\alpha$}
        \psfrag{E2}[r][r]{\footnotesize $- i\alpha$}
        \psfrag{E3}[lt][lt]{\footnotesize $2\pi + i\alpha$}
        \psfrag{E4}[lt][lt]{\footnotesize $2\pi -i\alpha$}
        \psfrag{2Pi}[b][t]{\footnotesize $2\pi$}
        \psfrag{Pi}[b][t]{\footnotesize $\pi$}
        \psfrag{0}[b][t]{\footnotesize $0$}
        \psfrag{U1}{$\fbox{1}$}
        \psfrag{U2}{$\fbox{2}$}
        \psfrag{U3}{$\fbox{3}$}
        \psfrag{U4}{$\fbox{4}$}
        \psfrag{L1}{$\fbox{1}$}
        \psfrag{L2}{$\fbox{2}$}
        \psfrag{L3}{$\fbox{3}$}
        \psfrag{L4}{$\fbox{4}$}
        \inputFigure{height=0.3\textheight}{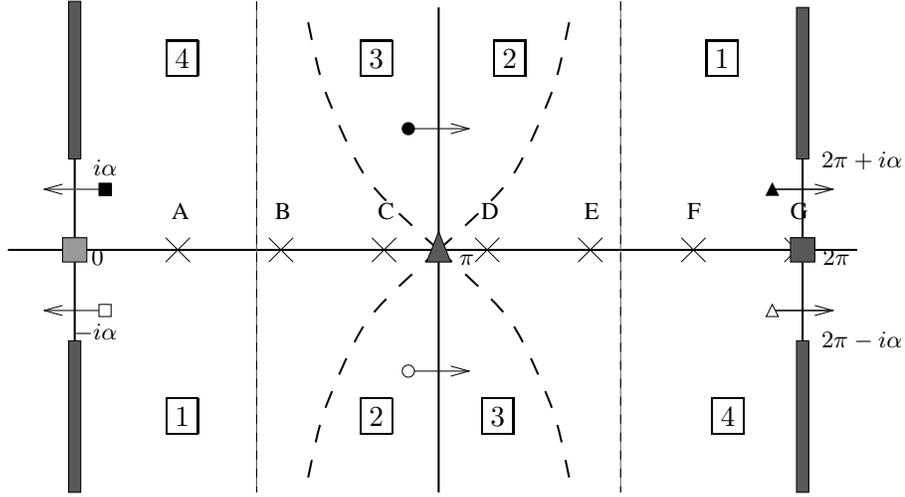}
        \protect\label{fig:wtot:w}
    \end{minipage}
  }\\
  \subfigure[$t$-- plane. 
    $v_c = i\protect\sqrt{z}(r_i + r_f - |r_f - r_i|)$
]{
    \begin{minipage}[b]{0.95\textwidth}
        \centering
        \psfrag{A}[][]{A}
        \psfrag{B}[][]{B}
        \psfrag{C}[][]{C}
        \psfrag{D}[][]{D}
        \psfrag{E}[][]{E}
        \psfrag{F}[][]{F}
        \psfrag{G}[][]{G}
        \psfrag{E1}[][]{-$\sqrt{i\eta}$}
        \psfrag{E2}[][]{-$\sqrt{v_c}$}
        \psfrag{E3}[][]{$\sqrt{v_c}$}
        \psfrag{E4}[][]{$\sqrt{i\eta}$}
        \inputFigure{height=0.3\textheight}{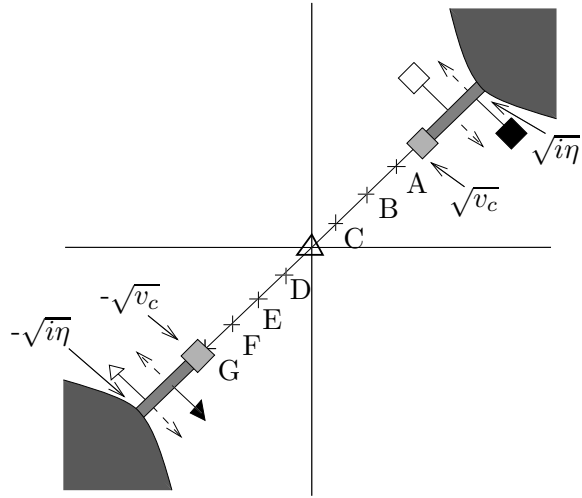}
        \protect\label{fig:wtot:t}
    \end{minipage}
  }
  \caption{$\omega\leftrightarrow t$ transformation: $n = 0$ branch. 
    Corresponding features in the two planes are labelled identically. 
    Labelled arrows are used here and in other figures to indicate 
    how branches are connected. 
  }
  \label{fig:wtot}
\end{figure}

\begin{figure}[p]
  \subfigure[$n=0$ branch]{
    \begin{minipage}[b]{0.48\textwidth}
        \centering
        \psfrag{Q1}{\footnotesize $\fbox{1}$}
        \psfrag{Q2}{\footnotesize $\fbox{2}$}
        \psfrag{Q3}{\footnotesize $\fbox{3}$}
        \psfrag{Q4}{\footnotesize $\fbox{4}$}
        \psfrag{1}{\footnotesize $1$}
        \psfrag{A}[][]{\footnotesize A}
        \psfrag{B}[][]{\footnotesize B}
        \psfrag{C}[][]{\footnotesize C}
        \psfrag{Uc}[][]{\footnotesize $u_c$}
        \inputFigure{scale=0.5}{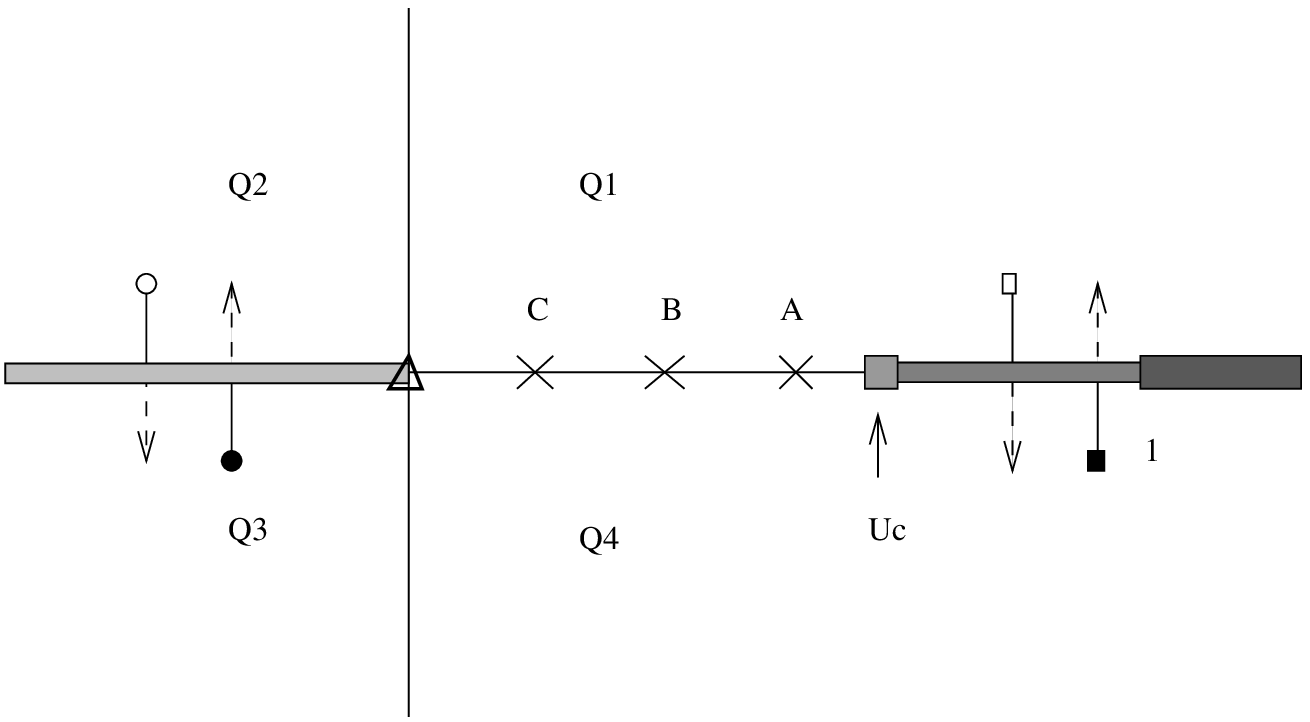}
        \protect\label{fig:u0}
    \end{minipage}
  }
  \subfigure[$n=1$ branch]{
    \begin{minipage}[b]{0.48\textwidth}
        \centering
        \psfrag{Q1}{\footnotesize $\fbox{1}$}
        \psfrag{Q2}{\footnotesize $\fbox{2}$}
        \psfrag{Q3}{\footnotesize $\fbox{3}$}
        \psfrag{Q4}{\footnotesize $\fbox{4}$}
        \psfrag{1}{\footnotesize $1$}
        \psfrag{D}[][]{\footnotesize D}
        \psfrag{E}[][]{\footnotesize E}
        \psfrag{F}[][]{\footnotesize F}
        \psfrag{G}[][]{\footnotesize G}
        \psfrag{Uc}[][]{\footnotesize $u_c$}
        \inputFigure{scale=0.5}{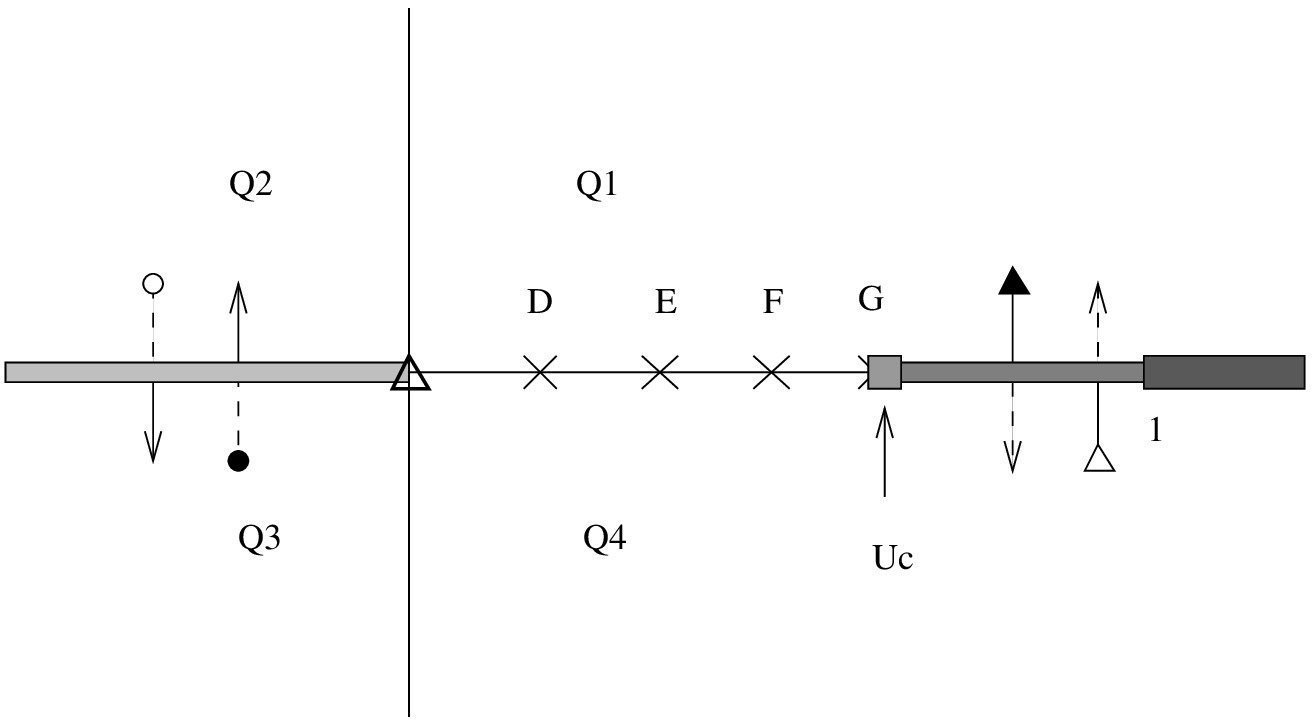}
        \protect\label{fig:u1}
    \end{minipage}
  }
    \caption{$\omega \leftrightarrow u \leftrightarrow v \leftrightarrow t$ 
    transformation: $u$ plane. $u_c = \frac{2}{\sigma}$}
    \protect\label{fig:u}
\end{figure}

\begin{figure}[p]
  \subfigure[$n=0$ branch]{
    \begin{minipage}[b]{0.48\textwidth}
        \centering
        \psfrag{A}[][]{\footnotesize A}
        \psfrag{B}[][]{\footnotesize B}
        \psfrag{C}[][]{\footnotesize C}
        \psfrag{iE}[][]{\footnotesize $i\eta$}
        \psfrag{Vc}[][]{\footnotesize $v_c$}
        \inputFigure{scale=0.45}{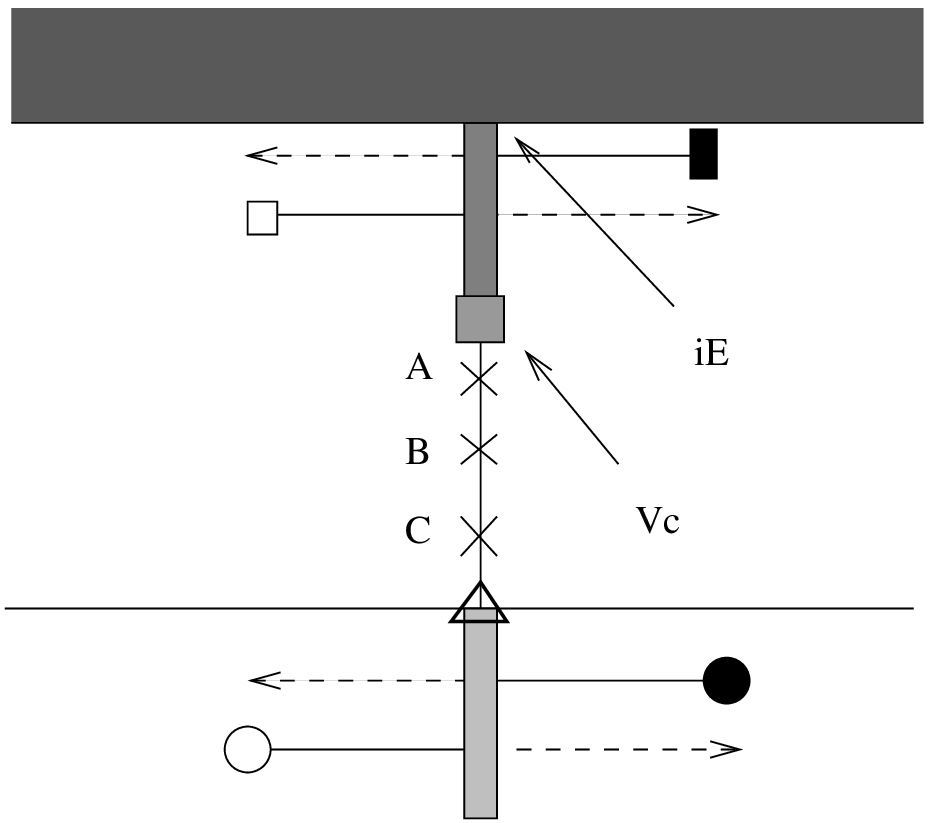}
        \protect\label{fig:v0}
    \end{minipage}
  }
  \subfigure[$n=1$ branch]{
    \begin{minipage}[b]{0.48\textwidth}
        \centering
        \psfrag{D}[][]{\footnotesize D}
        \psfrag{E}[][]{\footnotesize E}
        \psfrag{F}[][]{\footnotesize F}
        \psfrag{G}[][]{\footnotesize G}
        \psfrag{iE}[][]{\footnotesize $i\eta$}
        \psfrag{Vc}[][]{\footnotesize $v_c$}
        \inputFigure{scale=0.45}{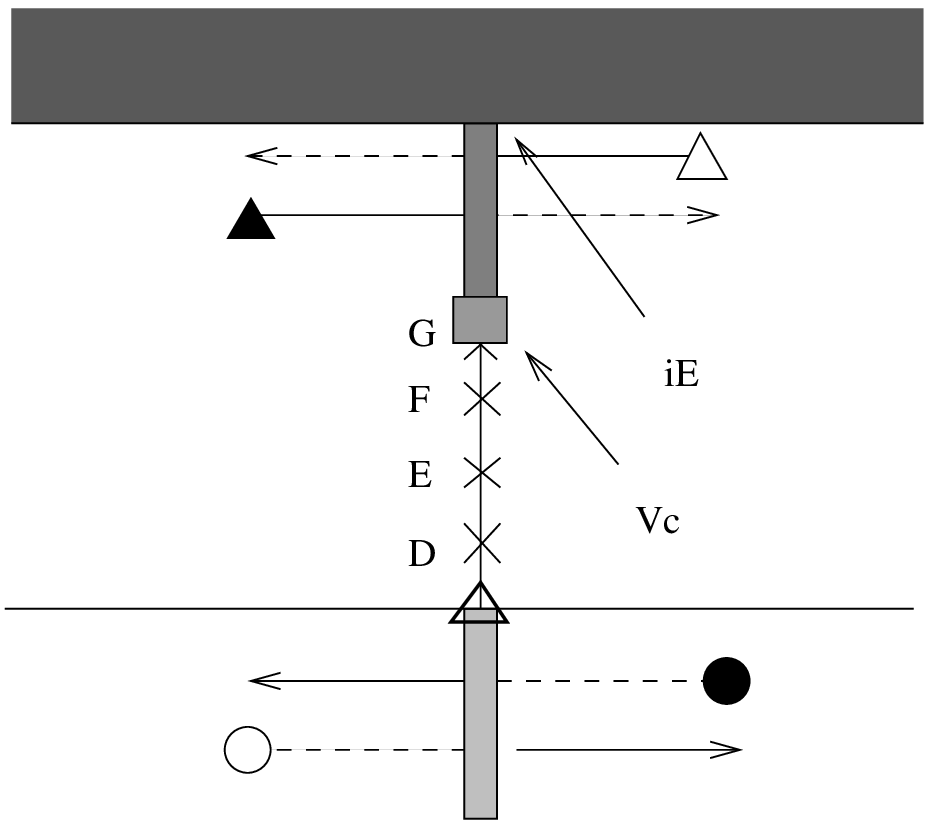}
        \protect\label{fig:v1}
    \end{minipage}
  }
    \caption{$\omega \leftrightarrow u \leftrightarrow v \leftrightarrow t$ 
    transformation: $v$ plane. 
    $v_c = i\protect\sqrt{z}(r_i + r_f - |r_f - r_i|)$.
    Analytic continuation into the darkly
    shaded regions corresponds to analytic continution through the cuts at
    $2n\pi \pm i\alpha, n\in\field{Z}$ in the $\omega$--plane.}
    \label{fig:v}
\end{figure}

\begin{figure}[p]
  \subfigure[$n=0$ branch]{
    \begin{minipage}[b]{0.48\textwidth}
        \centering
        \psfrag{A}[][]{\footnotesize A}
        \psfrag{B}[][]{\footnotesize B}
        \psfrag{C}[][]{\footnotesize C}
        \psfrag{iE}[][]{\footnotesize $\sqrt{i\eta}$}
        \psfrag{Vc}[][]{\footnotesize $\sqrt{v_c}$}
        \inputFigure{scale=0.45}{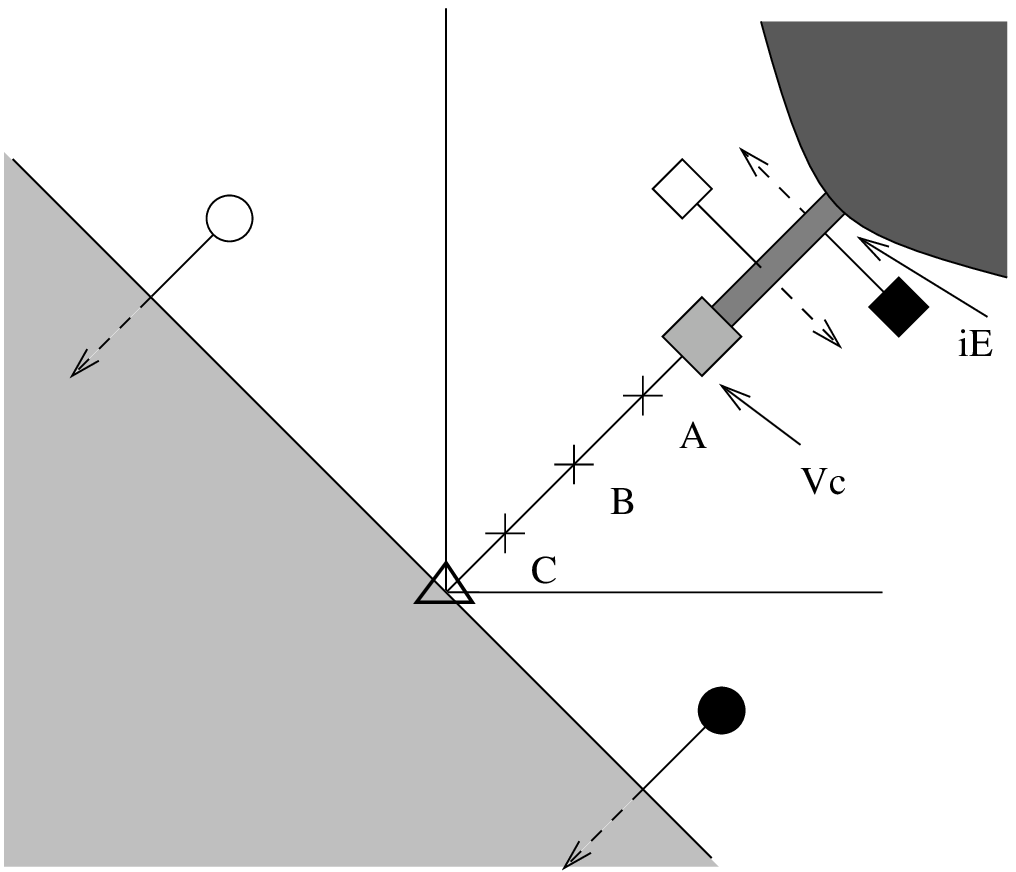}
        \protect\label{fig:t0}
    \end{minipage}
  }
  \subfigure[$n=1$ branch]{
    \begin{minipage}[b]{0.48\textwidth}
        \centering
        \psfrag{D}[][]{\footnotesize D}
        \psfrag{E}[][]{\footnotesize E}
        \psfrag{F}[][]{\footnotesize F}
        \psfrag{G}[][]{\footnotesize G}
        \psfrag{iE}[][]{\footnotesize $-\sqrt{i\eta}$}
        \psfrag{Vc}[][]{\footnotesize $-\sqrt{v_c}$}
        \inputFigure{scale=0.45}{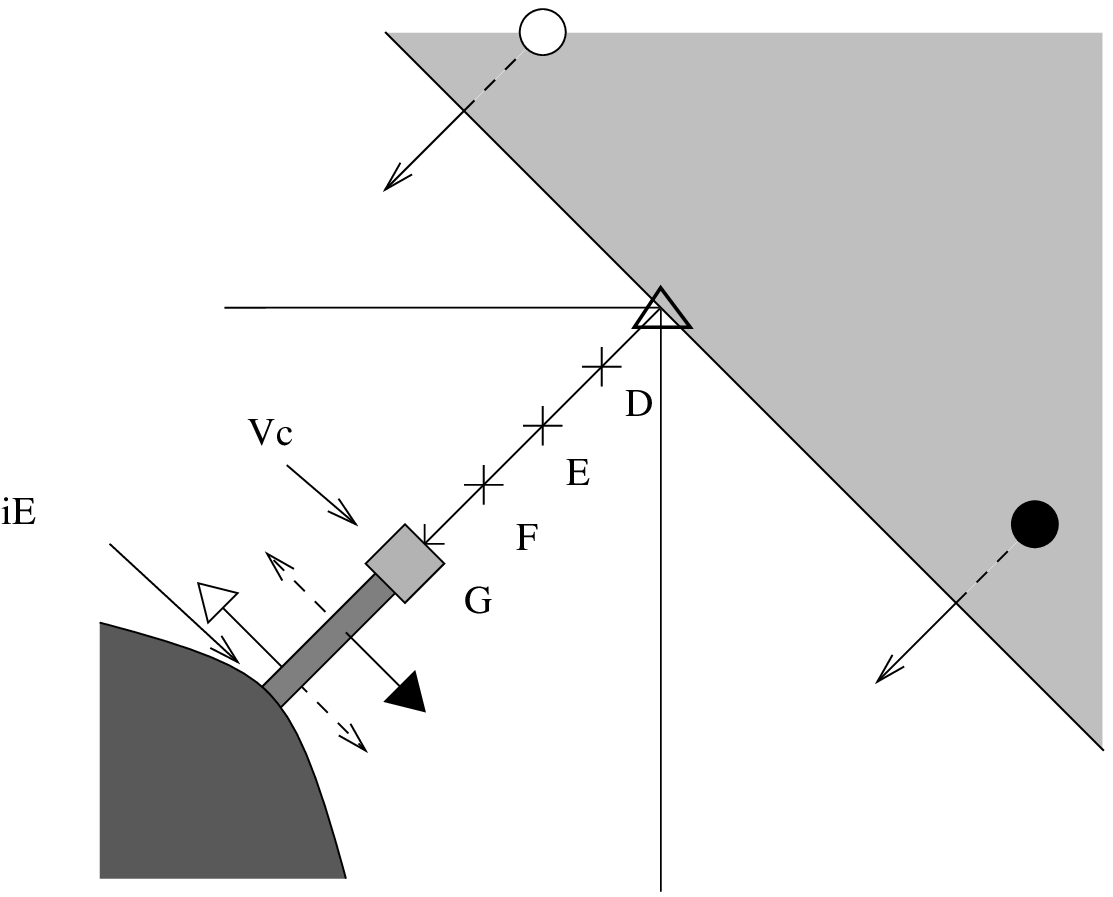}
        \protect\label{fig:t1}
    \end{minipage}
  }
    \caption{$\omega \leftrightarrow u \leftrightarrow v \leftrightarrow t$ 
    transformation: $t$ plane}
    \label{fig:t}
\end{figure}

$\omega\leftrightarrow t$ is summarised in Figure~\ref{fig:wtot}. The 
sequence of transformations leading to it is described in 
Figures~\ref{fig:wtot:w}, \ref{fig:u}, \ref{fig:v} and \ref{fig:t}.  The 
transformation is visualised by following features in the 
$\omega$ plane through the transformation sequence. It is also 
convenient to follow the quadrants of the $u$ plane.  
Arrows are used to denote how various branches should be glued 
together.
For example 
in Figure~\ref{fig:wtot:w} we have a arrow labelled with a solid 
circle which crosses the line $\Im(\omega) = \pi, \Re(\omega) > 0$.  In 
Figure~\ref{fig:u} we show the corresponding motion between branches 
of $u(\omega)$.  

Let us first consider $\omega \leftrightarrow u$. 
The images of the $u$ quadrants in the $\omega$ 
plane are determined in the following manner.  The images of the real 
and imaginary $u$ axes are readily determined. 
They partition the $\omega$--plane. One can identify which region 
corresponds to which quadrant of the $u$--plane by using the 
asymptotic result 
$u = 1 + \cos\omega \sim 1 + \tfrac{1}{2}e^{\mp i \Re(\omega)}$ as
$\Im(\omega) \to \pm\infty$ and continuity.
The transformation is clearly multivalued.  
Single valued transformations (branches) $\omega(u)$ can be obtained by
restricting the domain of $\omega$.  This can be done in an infinite 
number of ways so that we have the luxury of choosing the most convenient.
For example we can partition the 
$\omega$ plane into the regions $n\pi <\Re(\omega) < (n+1)\pi$,
$n \in\field{Z}$.
Each region maps into the $u$--plane in a $1\leftrightarrow1$ 
manner.  
Alternatively could choose the partition
 $2n\pi < \Re(\omega) < 2(n+1)\pi$, $\Im(\omega) >0$ and
$2n\pi < \Re(\omega) < 2(n+1)\pi$, $\Im(\omega) < 0$,
$n\in\field{Z}$.
Any set of curves can be used in to partition 
the $\omega$ plane provided the regions obtained each cover the 
entire $u$ plane once.  The convenient partition for our purposes is 
the first choice. 
In Figures~\ref{fig:wtot:w} and \ref{fig:u} we show the transformation 
for the $n=0$ and $n=1$ branches.  It will become apparent in due 
course that we need to follow two branches at a time.  For the moment 
the reader should note that the steepest descent contour through $\pi$ 
occupies both branches.


$u\leftrightarrow v$ is readily constucted using a sequence of 
elementary transformations .  The 
$v$-plane is given in Figure~\ref{fig:v}.  An important feature to 
follow is the branch cut at $u = \frac{2}{\sigma}$.  In the $v$--plane 
this becomes a vertical branch cut with the branch point at
\begin{equation}
	v_{c} = \left\{ 
		\begin{array}{ll}
             2i\sqrt{z} r_{i} & \mbox{for $r_{f} > r_{i}$}\\		
	         2i\sqrt{z} r_{f} & \mbox{for $r_{i} > r_{f}$}
		\end{array}
			\right.
\end{equation}

All that remains is $v \leftrightarrow t$. Although this is just a 
square root, some thought is necessary. Up until now we have used 
the symbol $\sqrt{\phantom{z}}$ with minimal explanation. In general 
this symbol denotes the standard square root, which has a branch cut along 
the negative real axis.
Let us define another square root whose branch cut is along the 
negative imaginary axis. As this can be a point of confusion 
let us be very precise
\begin{equation}
	\text{Let }
		z = |z|e^{i\theta}, 
             \text{ with } -\frac{\pi}{2} < \theta < \frac{3\pi}{2}.
    \text{ Then}
    	\left.\sqrt{z}\right|_{-i\infty} 
		= |z|^{1/2}e^{i\theta/2},
\end{equation}

We map the $n=0$ copy of the $v$--plane to the $t$--plane using $v = 
\left.\sqrt{z}\right|_{-i\infty}$ and the $n=1$ copy of the $v$--plane 
to the $t$--plane using $v = - \left.\sqrt{z}\right|_{-i\infty}$.  We 
show the results in Figure~\ref{fig:t} where it is clear that we can 
paste the two regions together.  The arrows indicate that this can be 
done in an analytic manner to give a $1\leftrightarrow1$ map of $0 
<\Re(\omega) < 2\pi \leftrightarrow$ (most of the) $t$--plane.  We 
denote this as $\omega_{0}(t)$ and call it the principal 
branch.

Let us then summarise algebraically.  We transform from $\omega$ 
to $t$ using
\begin{subequations}\label{eq:wtot}
\begin{gather}
	u  =  \frac{1 + \cos\omega}{\sigma}
	\\
	v  =  -i\eta( - 1 + \sqrt{1 - u})
	\\
	t = \begin{cases}
		   \phantom{-}\left.\sqrt{v}\right|_{-i\infty}
		   &  0 < \Re(\omega) < \pi \\		
	       -\left.\sqrt{v}\right|_{-i\infty}
	       & \pi < \Re(\omega) < 2\pi 
		\end{cases}
\end{gather}
\end{subequations}
and from $t$ to $\omega$ using
\begin{subequations}
\begin{gather}
	v  =  t^{2}
	\\
	u  =  1 - 
	 			\left( 1 + \frac{iv}{\eta}
	 			\right)^{2}
	\\
	\omega = 
		\begin{cases}
			\arccos(u\sigma - 1) 
			 &
			 	- \tfrac{\pi}{4} < \arg(t) < \tfrac{3\pi}{4} \\		
	       2\pi - \arccos(u\sigma - 1) 
	       & 
	          	- \tfrac{5\pi}{4} < \arg(t) < - \tfrac{\pi}{4}
		\end{cases}
\end{gather}
\end{subequations}
where we have used the principal branch of $\omega = 
\arccos(z)$.
This maps the $z$--plane into the region $0 <\Re(\omega) < \pi$.
The other branches of the partition are obtained in a similar manner.  
In fact one can readily see that $w_{n}(t) = 2n\pi + w_{0}(t)$, 
$n\in\field{Z}$. The steepest descent contour occupying $0 < \Re(\omega) < 
2\pi$ (with orientation given in 
Figure~\ref{fig:WedgeContourOmega}) is mapped to the path in the 
$t$--plane along the real axis from $\infty$ to $-\infty$. 

We can now see that it is not just the poles which spoil the Gaussian 
form of the integrand.  When $\min(|zr_{f}|,|zr_{}|) \ll 1$ (i.e.  
$\sigma \gg 1$) the cut at $\sqrt{v_{c}}$ collapses into the origin 
spoiling the Gaussian form of the integrand.  This is the origin of 
the more subtle, non--semiclassical behaviour previously 
mentioned.  At first sight it might 
appear that our formalism is always semiclassical, with its natural 
cutoff guaranteeing that $r_{f} + r_{i}$ has a minimum value of 
macroscopic magnitude.  However because $r_{f}$ and $r_{i}$ are not so 
bounded from below there remain benign but nevertheless fully quantum 
contributions in the corners.

\subsection{Uniform calculation of integral}

Deforming $C$ to the steepest descent contour and changing integration
variables to $t$, we readily obtain
\begin{equation}
	\textrm{Pole Terms} -
		K^{(0)}(r_{f},r_{i},\theta_{i}) +
		K^{(-1)}(r_{f},r_{i},\theta_{i})
\end{equation}
where 
\begin{equation}
	K^{(n)}(r_{f},r_{i},\theta_{i}) =
		\int_{-\infty}^{\infty} 
			P^{(n)}(t)
			f(t) e^{-t^{2}}
			dt 			
	\label{eq:KIntegrals}
\end{equation}
and
\begin{subequations}
\begin{align}
	P^{(n)}(t) & =  P(\theta_{i},\phi,\omega_{n}(t))
	\\
	f(t) & =  - \frac{1}{2\phi r_{f}}
				 t H_{1}^{(1)}(\eta + i t^{2}) e^{t^{2}} 
\end{align}
\end{subequations}
The integrations \eqref{eq:KIntegrals} are now in standard Gaussian 
form.

For our purposes the most convenient way to proceed is to evaluate the
integrals numerically using Gauss--Hermite quadrature. The quadrature
converges well if $f(t)P(t)$ is a smooth, slowly varying function.
Because our construction guarantees that $\eta$ is reasonably large, this is
true for $f(t)$. $P(t)$ however has poles and this condition doesn't apply
when the poles are close to the real axis. This corresponds to $\vec{x}_f$
being in the penumbral region~\cite{SmilanskyPenumbral} so that the Gaussian
approximation breaks down (see also \cite[Figure 11.9]{BornWolf80}).
We found that the presence of ``close'' poles spoiled the quadrature to an
unacceptable degree for a significant number of matrix elements. It thus
seems neccessary to handle such cases separately using some an alternative
numerical method. Fortunately, as we have mentioned, this is essentially a
Stokes' phenomenon so that we are able to isolate the penumbral behaviour
using well known uniform
approaches~\cite{BerryStokes91,Jones72,Bleistein,Pauli38}.

The basic idea is to analytically subtract the poles out of $P^{(i)}(t)$.
This can be done a number of ways, the simplest being a recursive
construction. 
Suppose we wish to subtract out poles at
$\{ \omega_{0}, \omega_{1}, \omega_{2},\ldots, \omega_{N}\}$. Let 
$t_{n} =t(\omega_{n})$ and form (temporarily dropping the superscript)
\begin{equation}
	Q_{0}(t) = P(t)(t - t_{0}) 
\end{equation}
Defining
\begin{subequations}
\begin{align}
	P_{0} & =  Q_{0}(t_{0})
	\\
	R_{0}(t) & =  \frac{Q_{0}(t) - P_{0}}{t - t_{0}}
\end{align}
\end{subequations}
we obtain
\begin{equation}
	P(t) = \frac{P_{0}}{t - t_{0}} + R_{0}(t)
\end{equation}
The pole at $t_0$ is thus isolated as a simple pole in the first term. We
can recursively subtract off other poles. Suppose that $N - 1$ poles
have already been subtracted off. Form
\begin{equation}
	Q_{N}(t) = R_{N-1}(t)(t - t_{N}) 
\end{equation}
and define
\begin{subequations}
\begin{align}
	P_{N} & = Q_{N}(t_{N})
	\\
	R_{N}(t) & =  \frac{Q_{N}(t) - P_{N}}{t - t_{N}}
\end{align}
\end{subequations}
we readily obtain
\begin{equation}\label{eq:PoleDecomposition}
	P(t) = \sum_{n=0}^{N} \frac{P_{n}}{t - t_{n}} + R_{N}(t)
\end{equation}
Although it is not immediately apparent from the construction,
$P_{n}$ is the residue of $P(t)$ at $t_{n}$. This is obvious from other
arguments in complex analysis. Evaluating the residue we get
\begin{equation}
	P_{n} = - \frac{\mu \phi}{4\pi} 
	\frac{\sin\omega_{n}}{\lambda(\omega_{n})t_{n}}
	\label{eq:residue}
\end{equation}

It is natural to order the poles according to their proximity to the 
real axis and to subtract them out in this order.  Plotting $R_n(t)$ 
after each subtraction we noted the smoothing effect.  As the 
proceedure is straightforward all the poles in the $t$--plane were 
subtracted out .

We now proceed with the uniform analysis of the wedge kernel. Deform the
Carslaw contour toward the steepest descent contour but this time don't
cross the poles. This gives two contours $T_{0}$ and $T_{-1}$ which are
totally contained in the $n=0$ and $n=-1$ branches of the $\omega$--plane.
Rewrite \eqref{eq:KIntegrals} as
\begin{equation}
		I^{(-1)}(r_{f},r_{i},\theta_{i}) - 
		I^{(0)}(r_{f},r_{i},\theta_{i})
\end{equation}
where
\begin{equation}
	I^{(n)}(r_{f},r_{i},\theta_{i}) =
		\int_{T_{i}} 
			P^{(n)}(t)
			f(t) e^{-t^{2}}
			dt 			
	\label{eq:IIntegrals}
\end{equation}
The image of $T_{0}$ $(T_{-1})$ in the $t$--plane is a contour
which goes from $-\infty$ to $\infty$ passing over (under) all the
poles under (over) the cut at 
$(-)\sqrt{v_c}$.

Using \eqref{eq:PoleDecomposition} and performing one last
one last pole subtraction we can write the integrand for $n=0$ as
\begin{equation}\label{eq:FullDecomposition}
  P^{(0)}(t) f(t) =
  \sum_{0 < \Re(\omega_{n}) < 2\pi}
  \frac{P^{(0)}_{n}f(t_{n})}{t - t_{n}} 
  +
  \sum_{0 < \Re(\omega_{n}) < 2\pi}
  P^{(0)}_{n}\frac{f(t) - f(t_{n})}{t - t_{n}}
  + R^{(0)}_{N}(t)
\end{equation}
with a similar result for $n=-1$.
The integration is analogously decomposed into three terms.
The absence of poles in the last two terms allows us
to deform the contour $T_{i}$ to the real axis. Using 
\begin{subequations}
\begin{align}
  \int_{T_{0}} 
  \frac{e^{-t^{2}}dt}{t - z}
  & =   \phantom{-} i\pi w(z)\phantom{-}\quad z\text{ above }T_{-1}
  \\
  \int_{T_{-1}} 
  \frac{e^{-t^{2}}dt}{t - z}
  & =   - i\pi w(-z)\quad z\text{ below }T_{0}
\end{align}
\end{subequations}
(Ref.~\cite{Abramowitz68}, Chapter 7) we finally obtain
\begin{equation}\label{eq:FullKernel}
\begin{split}
  K(r_{f},r_{i},
  &\theta_{i};\phi) =
  \\
  & \sum_{0 < \Re(\omega_{n}) < 2\pi} 
  i\pi P^{(0)}_{n}f(t_{n}) w(-t_{n})
  \quad +
  \sum_{-2\pi < \Re(\omega_{n}) < 0} 
  i\pi P^{(-1)}_{n}f(t_{n}) w(t_{n})
  \\
  & - \sum_{0 < \Re(\omega_{n}) < 2\pi}
  P^{(0)}_{n}\int_{-\infty}^{\infty}
  \frac{f(t) - f(t_{n})}{t - t_{n}} e^{-t^{2}}\,dt
  \\
  & +  
  \sum_{-2\pi < \Re(\omega_{n}) < 0}
  P^{(-1)}_{n}\int_{-\infty}^{\infty}
  \frac{f(t) - f(t_{n})}{t - t_{n}} e^{-t^{2}}\,dt
  \\ 
  & + \int_{-\infty}^{\infty} 
  \left( R^{(-1)}_{N^{(-1)}}(t) -   R^{(0)}_{N^{(0)}}(t)
  \right) e^{-t^{2}} \, dt		 
\end{split}
\end{equation}
where $N^{(0)}$ ($N^{(-1)}$) is the number of poles in the $n=0$ ($n = 
-1$) $t$--plane.

\eqref{eq:FullKernel} is quite an interesting form.  In the first two 
sums the contributions from the poles satisfying $-\pi 
<\Re(\omega_{n}) < \pi$ are the geometrical optics terms, modulated by 
an error function.  In this regime the asymptotic expansion of the 
error function has an exponential term which combines with the other 
factors to give the standard geometrical optics term.  The first two 
sums also contain other poles.  In the region containing these poles 
the exponential term is absent from the asymptotic expansion of the 
error function due to Stokes' phenomenon and the resulting terms in 
the sum are of higher order.  We shall refer to the first two sums as 
the \emph{uniform geometric} contribution to the wedge 
kernel.

The ``pole near a contour'' problem is now isolated in the (well 
documented) error function.  The remaining terms contain relatively 
smooth integrands and can be evaluated in a straightforward manner 
using Gauss--Hermite quadrature. They can also be evaluated 
asymptotically by expanding the integrands around the origin.

\emph{A priori}, one would now expect to have to consider the case
$\min(|zr_{f}|,|zr_{}|) \ll 1$. As is clear from Figure~\ref{fig:wtot:t},
the semiclassical approach is not suitable in this ``quantum'' regime 
in which
one should use the standard eigenfunction series. 
In Section~\ref{sec:Numerics} it will turn
out that the parameter $\min(|zr_{f}|,|zr_{}|)$ never gets small 
enough to require the use of this series.

\section{Validation: Numerical Results}
\label{sec:Numerics}

\subsection{Numerical discretisation}

The solution of integral equations via quadrature is a standard 
numerical technique described in many textbooks and with a vast 
research literature.  The basic idea is to first choose some quadrature 
rule to approximate the integration.  From here one readily obtains an 
inhomogenous linear equation for the unknown function at the 
quadrature nodes. This equation can then be solved using standard 
methods of numerical linear algebra. One then increases the number of 
nodes to achieve sufficient convergence.

In this section our main purpose is to validate the wedge--based 
approach.  To do this we content ourselves with the determination of
the first dozen or so eigenvalues for some typical triangles.  This 
can be done comfortably using a quadrature similar to that used by 
Berry and Wilkinson~\cite{BerryWilkinson} and carrying out the 
calculation entirely within the 
\emph{Mathematica} package~\cite{Mma}. Most cases could be done with an 
overnight run on a standard workstation.

We expect that one should be able to use more sophisticated approaches 
very profitably in the wedge--based case. We don't expect this to be 
so for the singular case.

\subsection{Quadrature}

The geometric nature of the integration variable, being an arc length 
along the boundary, allows one to readily visualise the quadrature.  
In order to keep things simple, we consider cases where the boundary 
is a uniform 1--dimensional lattice.  The length of each side is given 
as an integer multiple of some of a lattice spacing $\epsilon$ and we 
give these lengths in the form of a triple, $\{n_{0},n_{1},n_{2}\}$.  
 For example, the Pythagorean 
triangle with sides $1,\frac{4}{3}$ and $\frac{5}{3}$ can be specified 
by the triple $\{3, 4, 5\}$ with a lattice spacing of $ \epsilon= 
\frac{1}{3}$.  This defines 12 quadrature nodes.  We take 
$\{n_{0},n_{1},n_{2}\}$ and $\epsilon$ to define a base quadrature.  
We refine this quadrature to $d_{f}\{n_{0},n_{1},n_{2}\}$ and 
$\epsilon/d_{f}$ where the \emph{discretisation factor} $d_{f}$ is 
taken through the values $2,3,4,\ldots$ in order to demonstrate 
convergence.  For the wedge--based case we need to specify a 
subdivision point for each side.  This is readily accomodated in an 
obvious manner with the notation 
$\{\{n_{0},n_{1}\},\{n_{2},n_{3}\},\{n_{4},n_{5}\}\}$.

The quadrature nodes are chosen to lie at the midpoint of each unit 
cell so that the simplest quadrature is the midpoint rule. Our 
initial instinct was to avoid using the corners as quadrature points.

\subsection{Computational effort}

In order for the quadrature to be a good approximation it is necessary that 
one chooses enough nodal points so that the oscillations in the kernel 
are well sampled.  The kernel oscillates with a wave number of the 
order $\sqrt{z} L$ where $L$ denotes a typical length scale for the 
triangle.  A successful quadrature then requires the total number of 
nodes $n_{t}$ to be much larger than $\sqrt{z} L$.  Li and 
Robnik~\cite{LiRobnik95} use a parameter $b = 2\pi 
n_{t}/(\sqrt{z} L)$ which gives the number of number of nodes within 
one de Broglie wavelength.  An effective discretisation requires 
$n_{t} \gg \sqrt{z} L$ or $b \gg 1$.

A discretisation of order $n_{t}$ requires $O(n_{t}^{2})$ operations 
to evaluate the kernel and then $O(n_{t}^{3})$ operations to evaluate the 
determinant.  In our work the Bessel function evaluations were the 
most expensive part of the calculation so that effort grew 
quadratically in $\sqrt{z}L$.  As higher energies are considered the 
effort will eventually cross over to the cubic regime.

\subsection{Results}

The eigenvalues are determined from the zeros of the Fredholm 
determinant, $\det(1 - K)$, which we approximate using the discrete 
form of the kernel defined by our quadrature.

We consider in detail two cases: the equilateral triangle and the 
Pythagorean 3-4-5 triangle.  The former case is an integrable billiard 
and the eigenvalues are known exactly.  The integrable cases are 
described in~\cite{BerryWilkinson}.  The latter case was chosen as a 
``typical'' non--integrable billiard. 

Let us first give an overview of the lower part of the spectrum for 
each case, presenting plots of the modulus of the Fredholm 
determinant for real energies.  The cases considered are given in 
Table~\ref{Table:CasesDone}.
\begin{table}[htb]
  \centering
  \begin{tabular}{|c||c|}
    \hline
    Abbreviation & Base Quadrature  \\
    \hline\hline
    free-equilateral & $\{4,4,4\},\epsilon = \frac{1}{4}$ \\
    \hline
    free-345  &  $\{3,4,5\},\epsilon = \frac{1}{3}$ \\
    \hline
    wedge-equilateral 
    & $\{\{2,2\},\{2,2\},\{2,2\}\},\epsilon = \frac{1}{4}$  \\
    \hline
    wedge-345  & $\{\{2,1\},\{2,2\},\{3,2\}\},\epsilon = \frac{1}{3}$ \\
    \hline
  \end{tabular}
  \caption{Base quadratures for cases considered together with a 
  shorthand description.}
  \protect\label{Table:CasesDone}
\end{table} 

\begin{figure}[p]
    \subfigure[Free--based Fredholm determinant]{
    \begin{minipage}[b]{0.95\textwidth}
        \centering
        \psfrag{df = 1}[Br][Br]{\footnotesize$d_f = 1$}
        \psfrag{df = 2}[Br][Br]{\footnotesize$d_f = 2$}
        \psfrag{df = 3}[Br][Br]{\footnotesize$d_f = 3$}
        \psfrag{|Determinant|}{\footnotesize $|$Determinant$|$}
        \psfrag{Scaled Energy}[t][b]{\footnotesize Scaled Energy, $E_s$}
        \inputFigure{height=0.36\textheight}{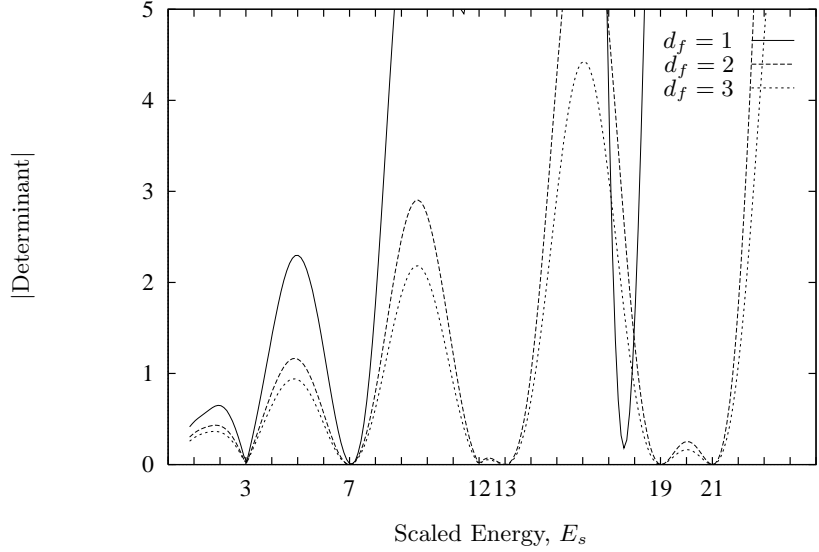}
        \protect\label{fig:Free444}
    \end{minipage}
    }\\
    \subfigure[Wedge--based Fredholm determinant]{
    \begin{minipage}[b]{0.95\textwidth}
	\centering 
        \psfrag{df = 1}[Br][Br]{\footnotesize$d_f = 1$}
        \psfrag{df = 2}[Br][Br]{\footnotesize$d_f = 2$}
        \psfrag{df = 3}[Br][Br]{\footnotesize$d_f = 3$}
        \psfrag{|Determinant|}{\footnotesize $|$Determinant$|$}
        \psfrag{Scaled Energy}[t][b]{\footnotesize Scaled Energy, $E_s$}
        \inputFigure{height=0.36\textheight}{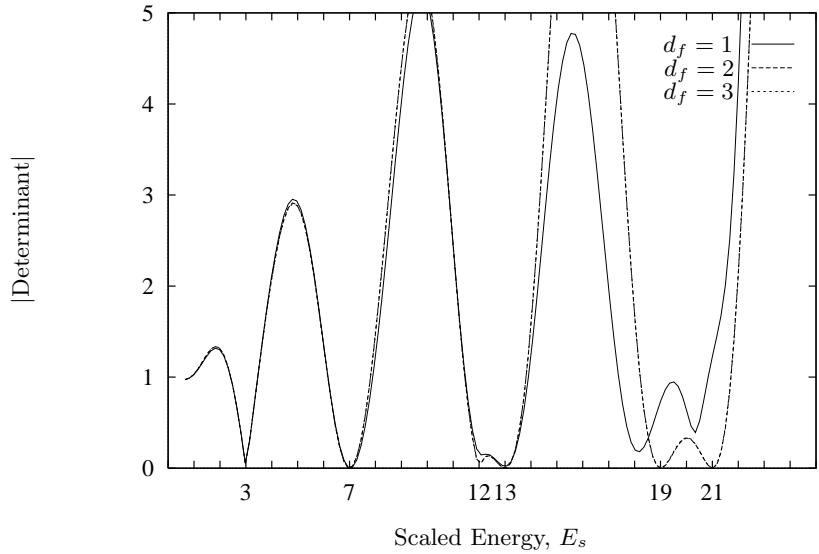} 
        \protect\label{fig:Wedge444} 
    \end{minipage}
    }
    \caption{Equilateral Triangle: solution of singular and non--singular
        integral equations by numerical discretisation.
	The energy has been scaled to
	facilitate easy comparision with the exact solution.  The
	number of lattice spacings per deBroglie wavelength, 
        $b = 6d_f/\protect\sqrt{E_s}$.  
	}
\end{figure}

In all plots we have applied an appropriate energy scaling.  For the 
integrable cases we have used a scaling which allows easy comparison
with exact results.  For the non--integrable cases we 
have used Weyl's result for the density of states to scale the energy 
so as to give, on average, a unit eigenvalue density.

In Figure~\ref{fig:Free444} we give the results for the 
free-equilateral case.  The first thing one notices is that the 
determinant does not converge with increasing $b$ i.e.  \emph{the 
continuum limit does not exist}.  
Despite this, the variation of the determinant still clearly resolves 
the eigenvalues.  The spectrum, given in appropriate units by $E = 
m^{2} + n^{2} - m n$, $1 < n < m $, is reproduced.  Cusp behaviour 
near the zeros indicates isolated eigenvalues and smooth behaviour 
indicates degeneracies.

The wedge-equilateral case is shown in Figure~\ref{fig:Wedge444}.  
Here one observes the convergence that 
Fredholm theory demands.  One expects that the lack of convergence in 
the free--based case is due to the singular nature of the integral 
equations.

\begin{figure}[p]
    \subfigure[Free--based Fredholm determinant]{
    \begin{minipage}[b]{0.95\textwidth}
        \centering
        \psfrag{df = 1}[Br][Br]{\footnotesize$d_f = 1$}
        \psfrag{df = 2}[Br][Br]{\footnotesize$d_f = 2$}
        \psfrag{df = 3}[Br][Br]{\footnotesize$d_f = 3$}
        \psfrag{|Determinant|}{\footnotesize $|$Determinant$|$}
        \psfrag{Scaled Energy}[t][b]{\footnotesize Scaled Energy, $E_s$}
        \inputFigure{height=0.36\textheight}{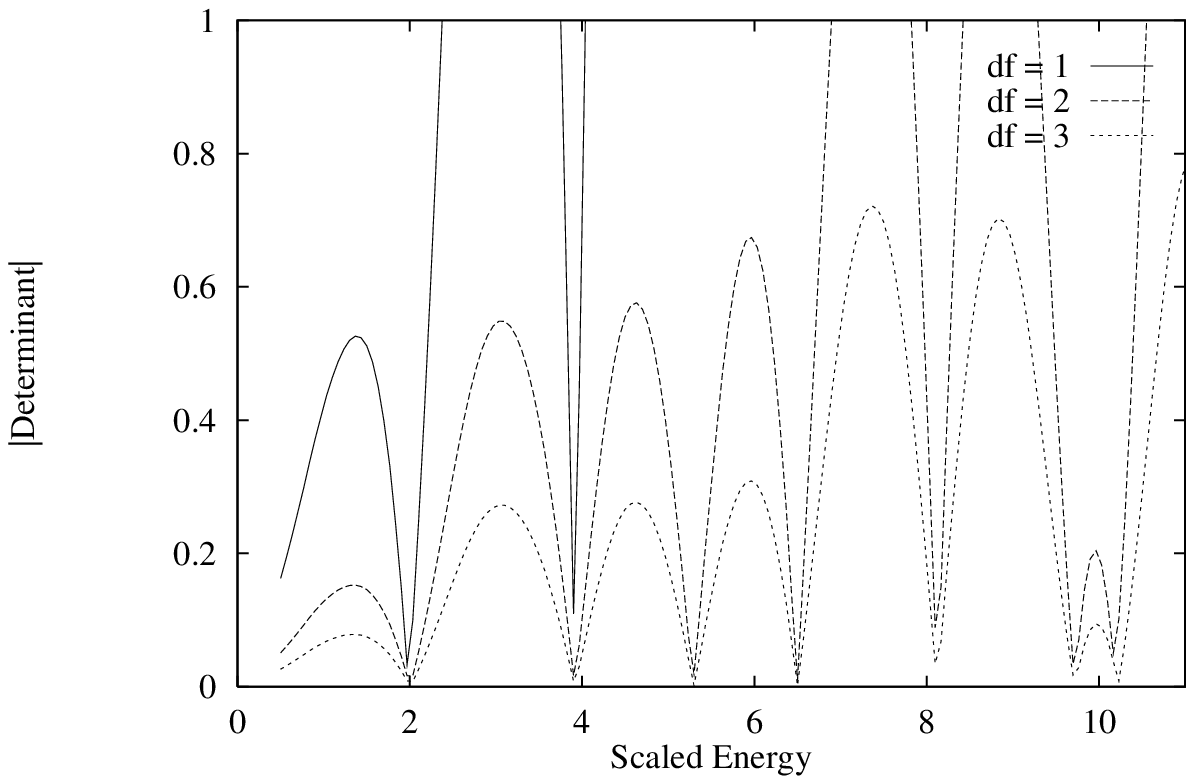}
        \protect\label{fig:Free345}
    \end{minipage}
    }\\
    \subfigure[Wedge--based Fredholm determinant: uniform geometric kernel]{
    \begin{minipage}[b]{0.95\textwidth}
	\centering 
        \psfrag{df = 1}[Br][Br]{\footnotesize$d_f = 1$}
        \psfrag{df = 2}[Br][Br]{\footnotesize$d_f = 2$}
        \psfrag{df = 3}[Br][Br]{\footnotesize$d_f = 3$}
        \psfrag{|Determinant|}{\footnotesize $|$Determinant$|$}
        \psfrag{Scaled Energy}[t][b]{\footnotesize Scaled Energy, $E_s$}
        \inputFigure{height=0.36\textheight}{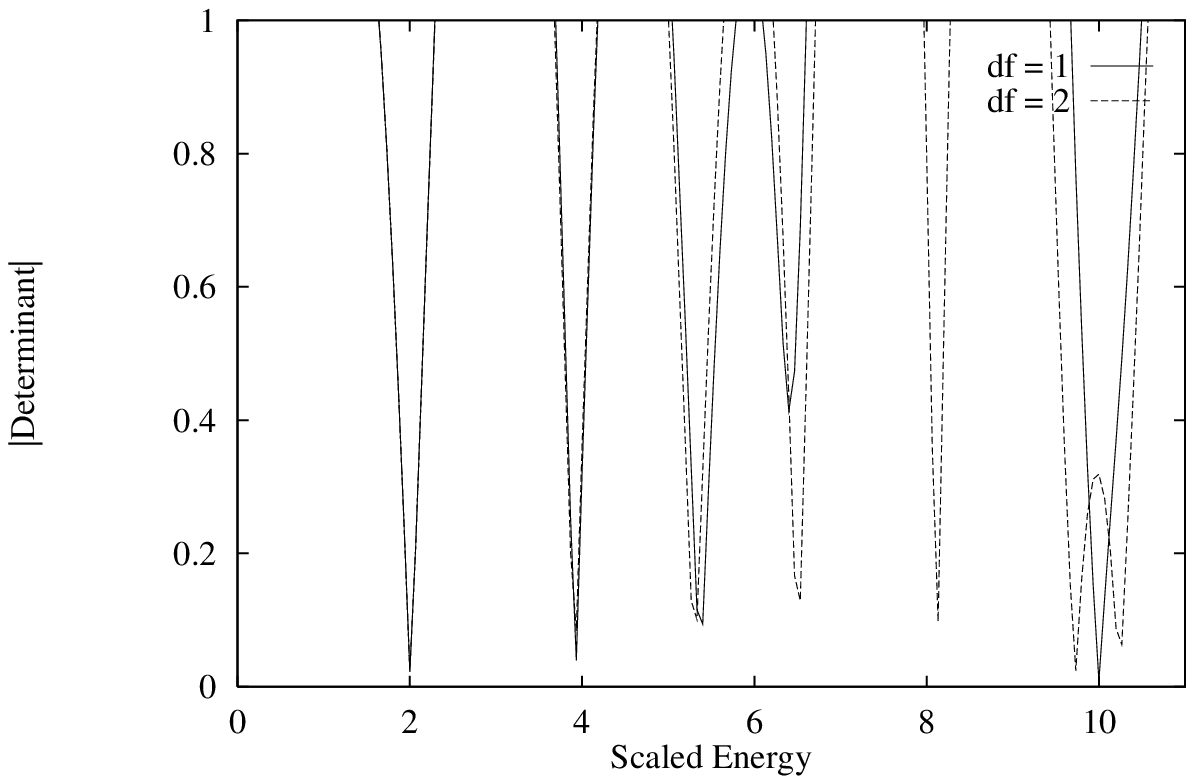} 
        \protect\label{fig:Wedge345} 
    \end{minipage}
    }
    \caption{3-4-5 Triangle: solution of singular and non--singular
        integral equations by numerical discretisation.
	The energy has been scaled to given a unit density of states.
	The number of lattice spacings per deBroglie wavelength, 
        $b = 4.3 d_f/\protect\sqrt{E_s}$.  
	}
\end{figure}

In Figures~\ref{fig:Free345} and ~\ref{fig:Wedge345} we show the 
corresponding results for the 3-4-5 triangle.  Once again  
the determinant doesn't converge in the free case. We have used the 
uniform--geometric component of the wedge kernel. As we shall see this 
gives accurate results.  

\begin{figure}[htb]
        \centering
        \psfrag{Energy2}[Br][Br]
            {\footnotesize $E_s = \phantom{1}2.0$}
        \psfrag{Energy6}[Br][Br]
            {\footnotesize $E_s = \phantom{1}6.0$}
        \psfrag{Energy10}[Br][Br]
            {\footnotesize $E_s = 10.0$}
        \psfrag{logdf}[t][b]{\footnotesize $\log d_f$}
        \psfrag{logDet}[][]{\footnotesize $\log |\det(d_f)|$}
        \inputFigure{height=0.3\textheight}{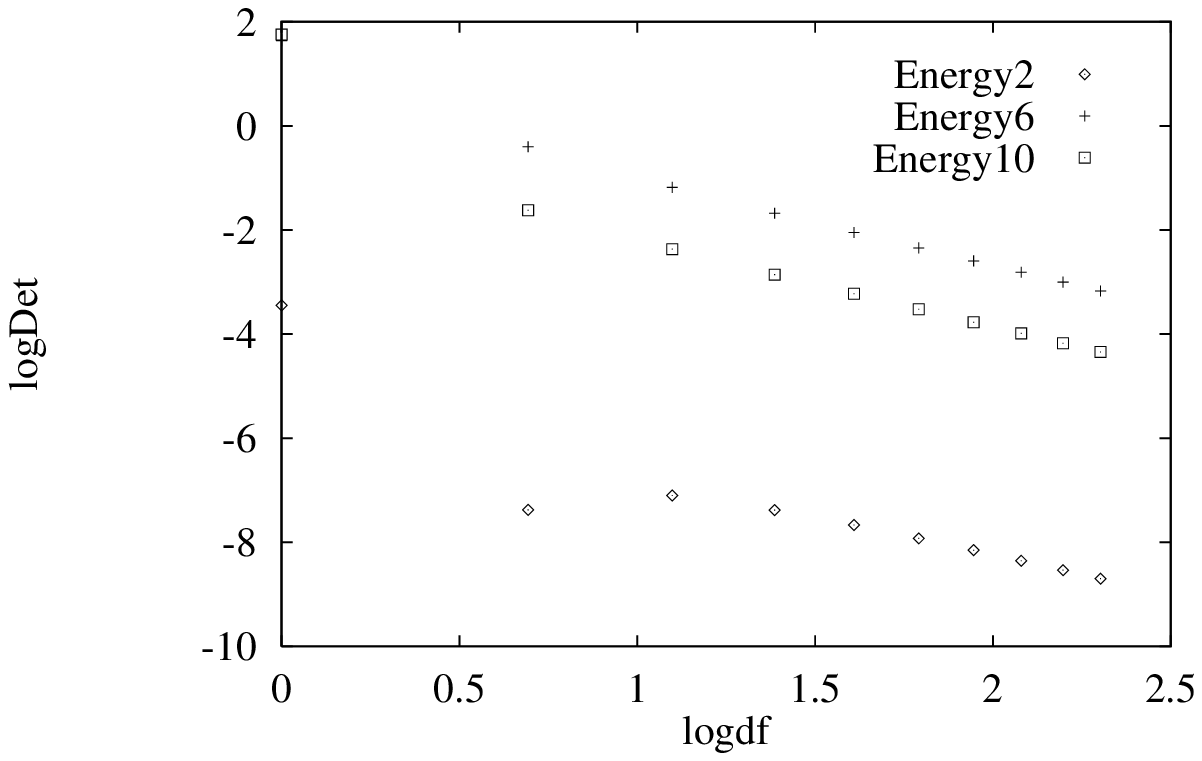}
	\caption{Convergence in free--based case: $\log-\log$ plot 
            of the modulus of the
            Fredholm determinant as a function of discretisation factor 
            $d_f$, $|\det(d_f)|$, at fixed energy for the 3-4-5 triangle.}
	\protect\label{fig:ConvFree345}
\end{figure}

In order to examine convergence more closely in the free--based 
approach, we show in 
Figure~\ref{fig:ConvFree345} a $\log-\log$ plot of the absolute value 
of the determinant versus the discretisation factor at some particular 
values of the energy for the 3-4-5 triangle.  It is clear that 
the determinant decreases with $d_{f}$ in a power law manner with a 
exponent $\approx -1.5$.  The exponent appears to be relatively uniform 
in energy.  One could thus arrest the lack of convergence with an 
\emph{ad--hoc} ``renormalisation'' factor.

\begin{figure}[htb]
        \centering
        \psfrag{Energy2.5}[Br][Br]{\footnotesize $E_s = \phantom{1}2.5$}
        \psfrag{Energy7.5}[Br][Br]{\footnotesize $E_s = \phantom{1}7.5$}
        \psfrag{Energy12.5}[Br][Br]{\footnotesize $E_s = 12.5$}
        \psfrag{logdf}[t][b]{\footnotesize $\log d_f$}
        \psfrag{logDiffDet}[][]%
            {\footnotesize $\log |\det(d_f+1) - \det(d_f)|$}
        \inputFigure{height=0.3\textheight}{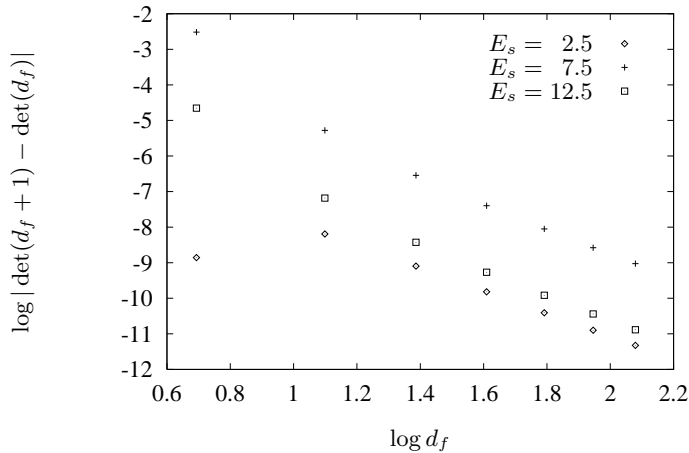}
	\caption{Convergence in wedge--based case: $\log-\log$ plot 
            of the first difference of the Fredholm determinant as 
            a function of discretisation factor 
            $d_f$, $|\det(d_f+1) - \det(d_f)|$, at fixed energy for 
            the 3-4-5 triangle.}
        \protect\label{fig:ConvWedge444}
\end{figure}

In Figure~\ref{fig:ConvWedge444} we quantify the convergence 
of the wedge--based approach by 
showing a $\log-\log$ plot of the first difference of the discretised 
Fredholm determinant versus $d_{f}$ for the equilateral 
triangle.  The power law exponent is approximatly -3, 
which is consistent with the use of the midpoint rule.  The 
straightness of the plot indicates that standard methods to 
accelerate convergence should work well.

\begin{figure}[htb]
        \centering
        \psfrag{Free df2}[Br][Br]{\footnotesize F, $d_f = 2$}
        \psfrag{Free df3}[Br][Br]{\footnotesize F, $d_f = 3$}
        \psfrag{Free df4}[Br][Br]{\footnotesize F, $d_f = 4$}
        \psfrag{WedgeGeom df2}[Br][Br]
                {\footnotesize WG, $d_f = 2$}
        \psfrag{WedgeUnifGeom df2}[Br][Br]
                {\footnotesize WU, $d_f = 2$}
        \psfrag{WedgeFull df2}[Br][Br]
                {\footnotesize W, $d_f = 2$}
        \psfrag{|Determinant|}{\footnotesize $|$Determinant$|$}
        \psfrag{Energy}[t][b]{\footnotesize Scaled Energy, $E_s$}
        \inputFigure{height=0.3\textheight}{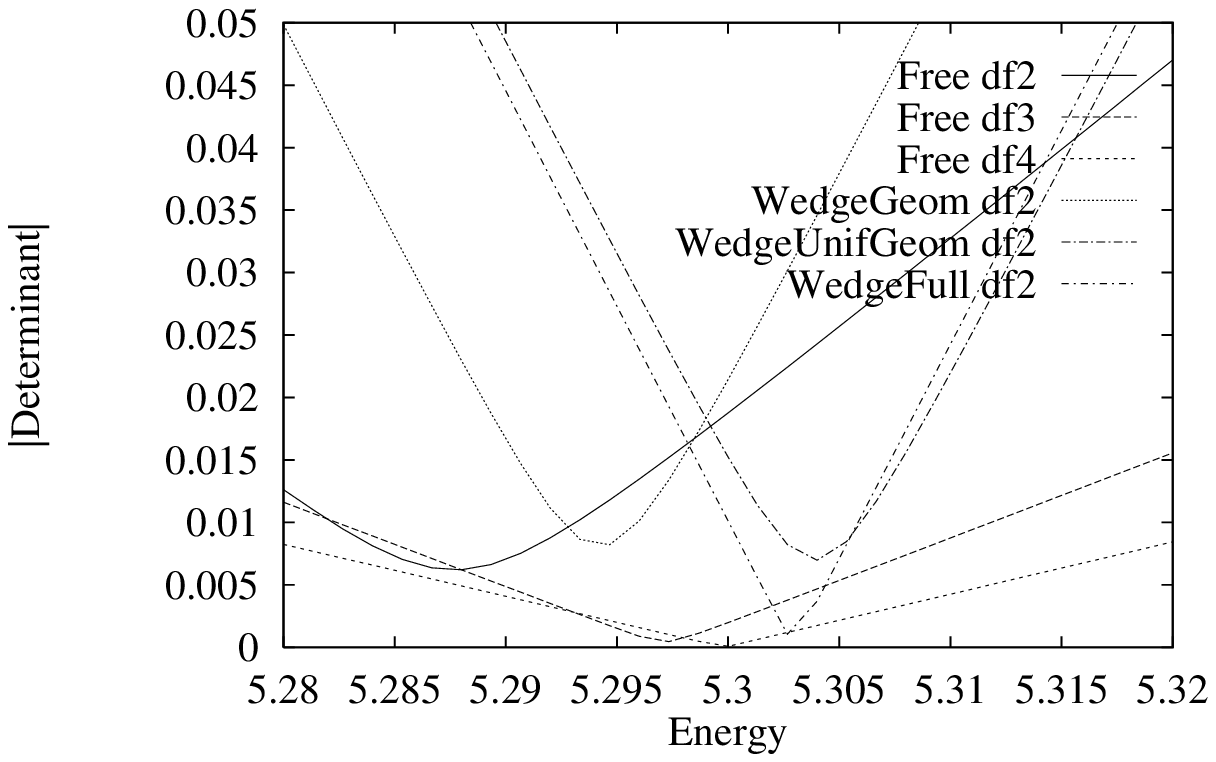}
	\caption{Fredholm determinants in the vicinity of the 3rd eigenvalue
            of the 3-4-5 triangle: detailed comparision of singular and 
            non--singular approaches. For the latter the effect of 
            approximations in the wedge kernel is also shown. F =
            free--based; WG = wedge--based, geometric kernel;
            WU = wedge--based, uniform geometric kernel;
            W = wedge-based with full kernel.
	}
	\protect\label{fig:Detail}
\end{figure}

In Figure~\ref{fig:Detail} we give detailed results in the vicinity of 
the 3rd eigenvalue of 3-4-5 triangle.  We compare the geometric, 
uniform--geometric and full kernel--based approximations for the 
wedge--based method together with 3 discretisations of the free--based 
result.  We see that the approximate wedge kernels give acceptable 
accuracy.  We also note that the wedge--based and free--based methods 
agree very well in the location of the 
zeros.

To obtain 6 figure accuracy uniformly across all matrix elements for 
the full kernel we used an 8-point quadrature for the Gaussian 
integrals in~\eqref{eq:FullKernel}. This is sufficient given the 
discretisation error in the numerical solution of the integral equation.
This quadrature is relatively expensive due to the need for extra
Bessel function evaluations.

As we mentioned at the end of Section~\ref{sec:Kernel}, the semiclassical
form of the wedge kernel is all we need. Because of the midpoint quadrature
there is a minimum value of $r_f$ and $r_i$. Our modest numerical goals
meant that the discretisation remained fairly coarse so that
$\min(|zr_{f}|,|zr_{}|)$ was never small enough to see the quantum regime.

In all figures one observes an expected decrease in the quality 
of eigenvalue determination as the quality of the discretisation $b$ 
decreases.  Where direct comparision is possible the wedge--based 
code seems to resolve the eigenvalues better for a given $b$.

In some of the figures the curves are not perfectly 
smooth.  This is due to the limited resolution of the plot.  Each plot 
displays the determinant sampled at about 100 equally spaced points.  
The samples are connected with lines by the plotting routine 
\texttt{gnuplot}. Where the determinant varies quickly, such 
as near an eigenvalue, a higher resolution is needed. It is best to 
offer a separate plot in such regimes, as we have done
in Figure~\ref{fig:Detail}. Such plots are needed to see the actual 
degree to which eigenvalues are resolved. For example in Figure
\ref{fig:Wedge345} the third eigenvalue does not appear to be 
resolved very well.  This is misleading because the the only energies 
sampled in this vicinity are $E_{s} = 5.267$ and $5.333$. 
Figure~\ref{fig:Detail} shows the true story.
Also the power law decrease of 
the free determinant with $d_f$ artificially smoothens plots. 
This might falsely lead one to think that the free--based code gives 
better resolution that the wedge--based code. A ``renormalisation'' of 
the free results would give a fairer comparision.

\subsection{Numerical checks}

In order to develop confidence in our results we applied several 
checks in order to diminish the possiblity that numerical or coding 
errors were present.  The first and most obvious check is the 
reproduction the exact spectrum for the equilateral triangle in both 
versions of the code. For the 3-4-5 case, and for nonintegrable triangles in general, there 
are no exact results for the eigenvalues.  In these cases we apply 
consistency checks.  Our main consistency check was to compare the 
results of the wedge--based code with the free--based code.  We found 
that for the common eigenvalues, the agreement was very good.  This 
can be seen in the case of the 3th eigenvalue in 
Figure~\ref{fig:Detail}.
As a further check we also considered the free--354 and free--534 
triangles, obtaining results identical to the free--345 case.  To 
eliminate the presence of integrable wedges, we also considered 
the 2-3-4 triangle.

The lack of determinant convergence for the free-based code was 
originally thought to be possibly a result of coding error.  The 
free--based code was originally intended as a ``warm--up'' exercise.  
After it was written we proceeded onto the more involved wedge-based 
calculation and improved the general quality of our software 
considerably.  It was then relatively easy to modify the wedge--based 
code to produce, independently, a new version of the free--based code, 
which reproduced the original results.  We further tested the 
free--based code by considering the other two integrable triangles: 
the isoceles right angled triangle (45,45,90) and the 
half--equilateral triangle (30,60,90).  Our present code is written 
for triangles with rational sides.  We thus considered the 
$\{12,12,17\}$ and $\{15,26,30\}$ triangles, with suitable values of 
$\epsilon$.  These approximate the two cases to 1 part in 100.  We 
found full agreement with the exact spectra.

\section{Periodic Orbit Expansions}
\label{sec:gtf}

We now have a mathematically well behaved reformulation of the 
boundary value problem which has been well validated using numerical 
techniques.

An intuitive understanding of the high energy, semiclassical 
eigenvalues could now be obtained by deriving a periodic orbit expansion 
using the Fredholm series, much as Georgeot and Prange 
describe~\cite{GeorgeotPrange95}.  The calculation of the periodic 
orbit expansions is now a matter of using the semiclassical form of 
the wedge kernel to calculate $\Tr_{n}$ asymptotically.  Similar 
analysis has been done in other 
contexts~\cite{PavloffSchmit95,AlonsoGaspard93}.  In our case the 
analysis will differ due to the subdivision of the sides of the 
polygon and the uniform approach to the diffractive contribution.

It is more important to first examine the nature of the periodic orbit 
expansions using Fredholm theory. In particular we wish to compare the 
computational effort of the periodic orbit expansions with  
standard numerical discretisation.

\subsection{Convergence of Fredholm series}

Standard Fredholm theory~\cite[\S 6.5]{Smithies62} shows that series for the 
Fredholm determinant,
\begin{equation}
        \Delta(z) = \sum_{n=0}^{\infty}\;\delta_{n}(z),
\end{equation}
is convergent by bounding $|\delta_{n}|$ using Hadamard's inequality.
From this bound one can readily estimate that $|\delta_{n}|$ 
increases until
\begin{equation}
	n_{c}  \approx \|K\|^{2}
\end{equation}
after which it decreases. In practice one might then expect that at 
least $3n_{c}$ terms are needed in the Fredholm expansion.

In order to apply this result we need to understand the behaviour of 
the norm \emph{for all values of the energy}.  In particular it is 
clear that one has \emph{ an effective semiclassical approximation if 
and only if the norm decreases with energy}.  The faster the rate of 
decrease the better. 

It is relatively easy to estimate the norm in the semiclassical limit
$\sqrt{z}|\vec{\Delta}_{\mu\nu}| \gg 1$. In this limit 
the wedge kernel is dominanted by geometric optics terms so that it 
has the form
\begin{equation} \label{eq:SemiClassicalNormEstimate}
	K_{\mu\nu}(t,s) 
		\sim
 		 - \frac{i}{\sqrt{2\pi}}\, z^{1/4} 
 			\sum_{\substack{
 				\text{\footnotesize classical}\\ 
 				\text{\footnotesize paths }\alpha}}
 			\frac{e^{i(\sqrt{z}|\vec{\Delta}^{\alpha}_{\mu\nu}| - 3\pi/4)
 			         }
 			     }{|\vec{\Delta}^{\alpha}_{\mu\nu}|^{1/2}}
 			\frac{\vec{\Delta}^{\alpha}_{\mu\nu}\,.\,\vec{n}_{\mu}}
 			{|\vec{\Delta}^{\alpha}_{\mu\nu}|}
\end{equation}
The number of classical paths varies with $t$ and $s$ as the shadow of 
the source point moves.  Diffractive contributions decay with a faster 
power law

The semiclassical norm then becomes
\begin{subequations}
\begin{align}
	\|K\|^{2}
	& \sim
	\sum_{\mu = 0}^{5} \sum_{\nu = 0}^{5}
		\int_{0}^{l_{\mu}}\,ds\int_{0}^{l_{\nu}}\,dt
			| K_{\mu\nu}(t,s) |^{2} 
	\\
    \begin{split}
    =
    \frac{\sqrt{z}}{2\pi}\;
    \sum_{\mu = 0}^{5} \sum_{\nu = 0}^{5}
		\int_{0}^{l_{\mu}}\,ds\int_{0}^{l_{\nu}}\,dt
	\left(
	\sum_{\alpha}
		\left(
		\frac{\vec{\Delta}^{\alpha}_{\mu\nu}\,.\,\vec{n}_{\mu}}
			{|\vec{\Delta}^{\alpha}_{\mu\nu}|}
		\right)^{2}
		\frac{1}{|\vec{\Delta}^{\alpha}_{\mu\nu}|}
	\right.
	\\
	\quad\quad
	\left.
	+ \quad 
	\sum_{\alpha \neq \beta}
		\frac{e^{i\sqrt{z}
			(|\vec{\Delta}^{\alpha}_{\mu\nu}| - 
			 |\vec{\Delta}^{\beta}_{\mu\nu}|
			)
	     }
		}{|\vec{\Delta}^{\alpha}_{\mu\nu}|^{1/2}
  		  |\vec{\Delta}^{\beta}_{\mu\nu}|^{1/2}
  		}
		\frac{\vec{\Delta}^{\alpha}_{\mu\nu}\,.\,\vec{n}_{\mu}}
			{|\vec{\Delta}^{\alpha}_{\mu\nu}|}
		\frac{\vec{\Delta}^{\beta}_{\mu\nu}\,.\,\vec{n}_{\mu}}
			{|\vec{\Delta}^{\beta}_{\mu\nu}|}
	\right)
	\end{split}
	\\
	&=
	\sqrt{z} L + \text{ oscillatory terms }
\end{align}
\end{subequations}
where $L$ is length of order the size of the triangle. The 
oscillatory terms come from a steepest descents evaluation of the 
integration over the second sum.

\eqref{eq:SemiClassicalNormEstimate} is not uniformly valid for all 
$t$ and $s$.  Although our technique avoids the unit source 
singularity there remains non--classical behaviour when one of 
$\sqrt{z}r_{i}$ or $\sqrt{z}r_{f}$ is very small.  One expects that 
\eqref{eq:SemiClassicalNormEstimate} will cross over into some other 
form in this boundary layer region.  We have no reason to believe that 
incorporating the boundary layer contribution will dramatically alter 
the above estimation of the norm.  To confirm this we generated 
$\log-\log$ plots of the norm versus energy and confirmed the above 
behaviour.  However our discretisation 
was not fine enough, nor was our energy high enough, to see any 
boundary layer behaviour in the calculation of the kernel.  It is thus 
conceivable that at higher energies some change in the norm behaviour 
may be observed.

We thus have 
\begin{equation}
		 n_{c} \sim \sqrt{z} L
\end{equation}
so that the number of terms needed grows with the square root of the energy. 
This accords with the literature on periodic orbit expansions but is 
not what one would expect of a semiclassical expansion.

\subsection{The correspondance principle}

In the old quantum theory Bohr presumed that any semiclassical 
approximation should improve in ``the limit of large quantum numbers''. In 
our context one would expect that the semiclassical approximation 
should improve as energy increased. This is clearly not the case for 
periodic orbit expansions.
The semiclassical Fredholm series is 
useful only if $\sqrt{z}L$ is \emph{large enough} to use the semiclassical 
approximation for $\delta_{n}(z)$ 
but \emph{small enough} so that not too many terms are needed.

We can offer an elementary analogy to impress apon the reader the true 
nature of periodic orbit expansions. 
Consider the asymptotic calculation of the zeros of the Bessel function,
i.e. solve $J_{\nu}(z) = 0$ for $z\to\infty$. 
The Bessel function can be defined in 
terms of series or integral representations. It is natural to handle 
this problem using the standard asymptotic result
\begin{equation} \label{eq:BesselAsymptotic}
	J_{\nu}(z) \sim \sqrt{\frac{2}{\pi z}} 
		\cos( z - \frac{1}{2}\nu\pi - \frac{1}{4}\pi)
\end{equation}
Zeros obtained in this way are very accurate and improve in accuracy 
as $z\to\infty$. In actual fact the accuracy is acceptable right down 
to the ground state! (\cite[9.5.12]{Abramowitz68}) 
A less natural way to proceed would be to use the series 
representation
\begin{equation} \label{eq:BesselSeries}
	J_{\nu}(z) = (\tfrac{1}{2}z)^{\nu}\,
		\sum_{k=0}^{\infty} 
			\frac{(-\frac{1}{4} z^{2}
			      )^{k}}
			     {k! \Gamma(\nu + k + 1)}
\end{equation}
because the number of terms need to obtain good convergence is of the 
order of $ \frac{3}{2} z $. 
Finding eigenvalues from periodic orbit expansions is like using the power 
series expansions to find zeros of Bessel functions. It is clear then 
that
 
\begin{quote}\itshape
	The problem of semiclassical quantisation of 
	chaotic systems will not be solved until one has found the analog of 
	\eqref{eq:BesselAsymptotic}.
\end{quote}

\subsection{Computational effort: comparison with discretisation}
\label{sec:GtfCompEffort}

Let us now examine the computational effort needed to evaluate the 
semiclassical Fredholm series.

The number of terms need to obtain good convergence is of order the 
dimensionless wavenumber $\sqrt{z} L$.  A semiclassical calculation of 
$\delta_{n}$ involves all periodic orbits with $n$ bounces for the 
triangular billiard. The number of such orbits is known to grow 
polynomially in $n$~\cite{Gutkin86}.  The order of this polynomial is not 
known but there are some conjectures.  For integrable triangles it is 
relatively easy to see that the growth is linear.  Either way the 
computational effort required to calculate high energy eigenvalues 
grows polynomially in $\sqrt{z} L$.  Discretisation of the 
boundary integral equation requires computational effort which grows 
like $(\sqrt{z} L)^{3}$ although in our case, as we discussed in 
Section~\ref{sec:Numerics}, it was effectively quadratic.  This 
becomes comparable to the semiclassical Fredholm series for 
\emph{integrable billiards}.  For generic billiards we would expect 
the effort to be qualitatively comparable

This is in contrast to the case of chaotic smooth billiards.  Our 
estimate of the norm in the semiclassical limit should essentially 
carry over to this case.  Following Georgeot and Prange it is clear 
that exponential proliferation causes the calculation of semiclassical 
Fredholm series to grow exponentially.  It is then clear that for 
smooth billiards the boundary integral method is much more efficient.
It thus appears that
\begin{quote}
\emph{Even with convergence 
questions settled, the problem of exponential proliferation remains. 
}
\end{quote}
We shall discuss this point further in 
Section~\ref{sec:discuss}.

\section{Discussion}
\label{sec:discuss}

\subsection{Singular integral equation}

\subsubsection{Analogy with quantum field theory}

The divergence of the norm in the free--based integral equation 
suggests that the Fredholm series itself may contain divergences.  In 
fact, using an argument virtually identical to the one we used for the 
norm, one can show that $Tr(K^2)$ diverges logarithmically.  One can 
extend this argument to show that the traces of all even powers of $K$ 
contain logarithmic divergences.  These divergences are contained in 
contributions which scatter back and forth between two given sides.  
It appears there are no other more serious divergences.  There is an 
obvious similarity between this behaviour and that of perturbation 
series in quantum field theory.

As in field theory, these divergences are due to the short distance 
behaviour of the theory.  
The semiclassical contribution comes from the long distance behaviour 
$\sqrt{z}L\gg 1$.  Because these two regimes are well separated it is 
easy to overlook the divergences in a semiclassical calculation.  
However, we know from the literature, in particular the work of 
Pavloff and Schmitt~\cite{PavloffSchmit95}, that periodic orbit 
expansions do give a reasonable degree of agreement.  This suggests 
that, like in field theory, the semiclassical properties may be 
insensitive to the short distance divergences.  Because of this we 
were tempted, at various stages of our study, to consider some sort of 
``renormalised'' periodic orbit expansion for the free--based 
approach.  The removal of singular behaviour in perturbation series is 
usually done in an \emph{ad--hoc} manner by ``regularising'' the 
divergences with artificially added convergence factors.  These depend 
on a cutoff parameter $1/\Lambda$ in such a way that 
$\Lambda\to\infty$ recovers the original integrand.  One evaluates 
the integrals for finite $\Lambda$, artificially removes the singular 
terms by some ``renormalisation'' proceedure, and then takes 
$\Lambda\to\infty$ to obtain ``finite'' results.

The natural regularisation for our problem is to introduce a short distance
cutoff, much as we do for the estimation of the norm with the definition of the
annular region. Alternatively we could apply a ``dimensional regularisation''
by replacing the $H^{(1)}_0$ in the free--Green's function with
$H^{(1)}_{\epsilon}$.

One then needs to develop a rationale to treat the divergent
parts of the traces, that is a ``renormalisation'' process. There is
nowhere obvious to ``absorb'' the divergences. Furthermore it is not clear
how this process would modify the periodic orbit expansion. Rather than
proceed along this line we developed the wedge--based approach.

We have no doubt that the pathologies in the singular formulation have
physical interpretations.  In particular we suspect that other boundary
value problems are mixed in as a result of the the sloppy treatment of the
corners. Perhaps this leads to a leaking of ``flux''.

\subsubsection{Wedge--based approach as renormalisation}

Our wedge--based construction clearly has the spirit of a ``regularisation''
because the division of each side into segments is clearly some sort of
``cutoff''. However our approach is not a \emph{post--facto} arbitrary
alteration of the result, it is a first principles \emph{ab--initio}
construction. 

The need for a wedge--based construction is also indicated by the nature of
the divergences in the singular approach. As we indicated the divergences
came from repeated propagation between two given sides. In the perturbation
series culture one would have removed these divergences by summing over
``ladder'' diagrams. We can interpret the perturbation about the wedge
solution as a way of summing over these diagrams. The Faddeev equations have
a similar interpretation.

Even though the wedge--based approach is clearly preferable there are
circumstances where such an approach is not available. Our study provides
a simple model in which one can critically examine the concept of
``renormalisation'' in more depth.

\subsubsection{Wedge--based approach and renormalisation group}

When we first came across the need to subdivide the sides of the triangle,
we were quite uncomfortable about the arbitrariness of the subdivision. We
have come to think that this subdivision cannot be avoided. In actual fact
there is a natural choice of subdivision: given the role of the norm in
controlling the convergence of the Fredholm series it is natural to minimise
the norm with respect to this subdivision. 

In many ways this is like the field theoretic renormalisation group program.
Here one introduces an arbitrary parameter, the renormalisation point $\mu$.
This is where one fixes coupling constants. Physical quantities should be
independent of $\mu$ and this is imposed in the theory by setting
the first derivative with respect to $\mu$ to zero. In our case setting the
derivative to zero is clearly associated with improving convergence.

\subsection{Smooth billiards}
\label{sec:discussSmoothBill}

\subsubsection{Doubts concerning the standard approach}

The work of Georgeot and Prange suggests that one can immediately obtain
semiclassical Fredholm series for smooth billiards from the integral
equations of Balian and Bloch. Our experience with the triangular billiard
has led us to entertain doubt and conjecture that these non--singular
integral equations are incorrect. As we hope to have impressed upon the
reader, the major conceptual difficulty in our work has been the
determination of mathematically well behaved integral equations.

We have two pieces of evidence for this conjecture.  First of all, a careful
boundary limit along the lines of Section~\ref{sec:FreeBased} seems to
produce $\epsilon$ dependence similar to that in the corner boundary layers
in the triangle. Secondly, there are problems in the standard approach for
non--convex domains.  Li and Robnik~\cite{LiRobnik95} have shown that the
boundary integral method fails.  Balian and Bloch have shown that
classically forbidden chordal paths in the periodic orbit expansion
cancel~\cite[II]{BalianBloch}.  In contrast there is no difficultly with
non--convex polygonal domains in our approach. Some matrix elements will
have no geometric contributions, but there will still be diffractive
contributions which creep around the corners much as one would
expect.  We thus feel that the kernel for smooth billiards should have a
diffractive component in order to accomodate the non--convex case.

We suspect the Balian and Bloch calculation is correct only for sufficiently
smoothed density of states expansions.  In fact this is all they claim, but
their work seems to be regarded in the same light as the Gutzwiller trace
formula, i.e. capable of determining individual eigenvalues.

We thus feel that a critical look at smooth billiards is needed. In 
particular we offer some alternative approaches.

\subsubsection{Alternative approaches to smooth billiards}

With suitable solutions to perturb about, our construction can be readily
generalised to chaotic billiards. Given a suitable form of the Green's
function outside a disk, one could readily obtain non--singular integral
equations for the 3--disk system.  Here the finite separation of the disks
gives a lower bound on propagation distances so that the resulting integral
equations will clearly be non--singular.  Given the Green's function for the
hyperbola wedge, which might be derived from eigenfunction expansions,
we could obtain convergent periodic orbit expansions for the hyperbola
billiard.

An interesting and more general proceedure would be to approach smooth
billiards using a suitable sequence of polygonal approximations. One could
take such a limit before or after the integral equations are solved.  A
natural starting point would be to consider the regular n-gon and take the
limit $n\to\infty$ to see what sort of integral equation one obtains for the
circle.  This approach has been used by Kac~\cite{KacDrum66} to derive 
the curvature dependent constant term in the Weyl formula for smooth 
billiards.  Before using this approch he tried to ``perturb'' about 
the solution of the diffusion equation on a disk.  He was forced into 
using a polygonal approach due to the lack of a ``workable'' solution 
to this problem.

\subsection{Modified periodic orbit expansions}

Whether or not the standard integral equations for smooth billiards need to
be replaced, much of our discussion is relevant to more general systems. Let
us compare then our approach with the various modifications of the original
Gutzwillers trace formula.

\subsubsection{Surface of section techniques}

The surface of section approach of Bogomolny~\cite{Bogomolny,
Lauritzen,Haggerty} takes a ``section'' in configuration space and expresses
the eigenvalues as the zeros of a Fredholm determinant $\det( I - T )$.  $T$
is an operator whose co--ordinate space representation is the semiclassical
propagator containing all classical paths starting and ending, but not
otherwise crossing, the surface of section.  This method is only partially
semiclassical since the Fredholm determinant has to be determined by a
quantum technique: either discretision or diagonalisation in some basis.
Bogomolny's technique is generally applicable and is not restricted to
billiard problems.  It was, however, motivated by the boundary integral
approach to billiard problems.

Our approach is a semi--rigorous realisation of Bogomolny's ideas. The
semiclassical expansion of the wedge propagator is clearly analogous to the
$T$ matrix. Bogolmony's approach corresponds to treating the wedge kernel
semiclassically while treating the Fredholm determinant fully quantum
mechanically. In our approach the Fredholm determinant arises as a natural
consequence of the reformulation of the problem as an integral equation.

As we discussed in Section~\ref{sec:GtfCompEffort} the surface of
section/boundary integral equation method requires much less computation
effort. The sacrifice of classical intuition in treating the Fredholm
determinant quantum mechanically seems a necessary price to pay in order to
avoid exponential proliferation.

While most of the literature views the problem of quantum chaos as
determining quantum mechanical quantities from classical ones, some authors
also considers the converse: studying classical mechanics using quantum
mechanics. This point of view arises from regarding the semiclassical
Fredholm series as a \emph{generating function for classical periodic
orbits}. One might regard quantum mechanics as a rival of the thermodynamic
formalism, which considers similar generating functions. One might even muse
that if experiment hadn't demanded it, one might have had to develop
quantum mechanics as a linearisation of classical mechanics. From this point
of view it appears that Bogomolny's hybrid method may be the most effective
way to study both quantum \emph{and} classical mechanics

\subsubsection{Cycle expansions}

As Georgeot and Prange have pointed out, the pseudo--orbit expansion, which
arises from the expansion of the zeta function, is the Fredholm series. In a
similar manner cycle expansions are also the result of the expansion of zeta
functions. The way in which terms are grouped together so that longer orbits
are grouped with shorter ones is very similar to the form of the Fredholm
series~\cite{BogoZeta92}.

Cycle expansions, however, have a further rationale within classical
mechanics~\cite{CvitanovicPhysica}. The basic element of cycle expansions is
that the dynamics is topologically organised in terms of a symbolic
dynamics. This involves a considered partition of the phase space. Each
region is labelled with a symbol so that each orbit, in particular each
periodic orbit, has a symbol sequence associated with it. This topological
understanding allows one to construct well behaved cycle expansions of the
Fredholm determinant. The cycle expansions themselves come from expansions
of determinants associated with automata describing the grammar of the
allowed symbols. Our wedge construction is similar in spirit in that the
phase space (the boundary) is divided into regions. In our case the form of
the division is not dictated by the classical dynamics but by the need to
generate non--singular integral equations. We feel that there should be a
relationship between these different criteria. In particular we feel that
the quantum mechanics of a problem might provide a useful guide to
developing suitable symbolic dynamics for the classical problem.

\subsection{Precursors of the Fredholm series}

Whilst the use of the Fredholm theory began with Georgeot and Prange, the
Fredholm series, as they point out, has already appeared in the guise of the
pseudo--orbit and cycle expansions.  It has thus been a part of the quantum
chaos literature for some time.  As another example Alsonso and Gaspard use
the $\log\det = \Tr\log$ identity~\cite[Eqn. (2.8)]{AlonsoGaspard93} in a
Balian--Bloch context. Bogolmony~\cite[Eqns (23) and (41)]{BogoZeta92} gives
arguments for the convergence of the Fredholm series in the context of his
surface of section technique without specifically refering to Fredholm
theory.

In much of the literature emphasis is given to the zeta function, probably
as a result of the influence of the number theoretical and hyperbolic
geometry literature. This influence then motivates the ``solution'' of the
convergence problem by formally expanding out the zeta function and grouping
the terms in a well motivated, but essentially \emph{ad--hoc} way.
\emph{The Fredholm theory, with its emphasis on the
integral equation reformulation of the boundary value problem, provides a
direct and mathematically firm way to circumvent this roundabout procedure}.

\subsection{Improving convergence?}

Exponential proliferation is only a problem in the semiclassical 
Fredholm series because the norm of the kernel grows like $\sqrt{z}L$.  
The question arises as to whether it is possible to further manipulate 
the integral equations to reduce this norm.  In the literature on 
integral equations it sometimes happens that iterating the integral 
equations can help.  This leads to the consideration of the Fredholm 
Determinant $\det(1 + K^n)$.  It is fairly clear, however, that the 
semiclassical form of $K^n$ is of similar character to that of $K$ so 
that the argument we presented for the norm should still apply.  While 
we are not hopeful that such manipulations will lead to a norm which 
decreases with energy we feel that it is best to defer a full 
semiclassical analysis until this issue is fully 
investigated.

\subsection{Deriving Keller's geometrical theory of diffraction}

Our wedge--based approach is an example of a way in which one may 
place Keller's geometrical theory of diffraction~\cite{KellerGTD62} on 
a mathematically firmer basis.  Keller's theory begins with the 
standard semiclassical solutions of the wave equation $Ae^{iS}$ in 
which the action $S$ is determined via an extended Fermat principle, 
which includes diffracted paths.  Because these are approximate 
solutions to a \emph{differential} equation they contain arbitrary 
constants.  These have to be determined using the boundary conditions.  
Keller does this by matching the approximate solutions to exactly 
known solutions near points of diffraction.  The idea here is that in 
the short wavelength limit only the local geometry can effect the 
solution.  Often the local geometry is one in which the problem has 
been exactly solved.  One then expects to be able to obtain the local 
behaviour of the solution as a perturbation of this exact result.  
Matching both solutions to leading order will give the unknown 
constants.  The spirit of Keller's technique is thus similar to that 
presented here.  However, because we use Green's theorem and convert 
the problem to an \emph{integral} equation, boundary conditions are 
automatically incorporated so that the somewhat awkward matching 
proceedure is not needed.  Furthermore ours is an exact reformulation 
of the problem and the semiclassical approximation is done 
\emph{after} the exact solution is obtained.

\section*{Acknowledgements}
\addcontentsline{toc}{section}{Acknowledgements}

The support of the Australian Research Council is gratefully 
acknowledged.

\appendix
\section{Wedge Greens Function}
\label{sec:wedgapp}

A statement of the boundary value problem for the wedge Green's function is
obtained from \eqref{eq:TriangleBVP} by simply replacing the domain
$\triangle$ with the wedge
$ W = \{ (r,\theta) : 0 < \theta < \phi, 0 < r < \infty\}$ and requiring
that the solution decay at infinity.

Carslaw~\cite{Carslaw18} derived this Green's function using a method due to
Sommerfeld~\cite{Carslaw99}. The solution is constructed by altering a
trivial contour integral representation for a suitable elementary solution.
While the motivation for this proceedure is somewhat obscure it is
reasonably easy to verify the result. Here we offer an outline of the
derivation which combines Carslaw's discussion with a method due to
Williams~\cite{Williams59}. Once again we shall try to maintain reasonable
standards of rigor, but our discussion will not be rigorous. The reader
can find yet another derivation in~\cite{Oberhettinger58}.

We begin following Williams with the free Greens function in polar
co--ordinates,
\begin{equation}
    G_0(r,r_{i};\theta - \theta_{i}) =
    	-\frac{i}{4}H_{0}^{(1)}
    	\left(
    	\sqrt{z^{\phantom{2}}}
    	\sqrt{r^{2} + r_{i}^{2} - 2 r r_{i} \cos(\theta - \theta_{i})}
    	\right)
    \label{eq:PolarFreeGF}
\end{equation}
We rewrite the Helmholtz equation which it satisfies in the following 
form
\begin{equation}
	\frac{1}{r} \frac{\partial}{\partial r}
		\left( r \frac{\partial G_0}{\partial r}
		\right) 
	+ z G_0
	= 
	- \frac{1}{r^{2}} \frac{\partial^{2}G_0}{d\omega^{2}}
	\label{eq:PolarHelmholtz}
\end{equation}
where we have set $\theta - \theta_{i} = \omega$, promoting it 
to a complex variable,

One then writes down a angular superposition of elementary solutions
as an ansatz 
\begin{equation}
	G_{W}(r,\theta) = \int_{C} 
		f(\omega,\theta) G_0(r,r_{i};\omega)\,d\omega
	\label{eq:ansatz}
\end{equation}
The contour $C$ is a suitably chosen path in the complex $\omega$ 
plane.  Note the similarity to the method of separation of variables.

In order to make practical use of this ansatz we must be able to interchange
integration, differentiation and limits.  Such manipulations are valid only
if integral is uniformly convergent. The contour $C$ must then avoid
singularities of the integrand by a distance which is bounded below.
Furthermore in any passage of the contour to infinity the integrand must go
to zero sufficiently fast.

We must also specify the branches of the square roots appearing in
\eqref{eq:PolarFreeGF}. This requires some anticipation of the 
ultimate form of the solution.  From integrable cases and 
semiclassical arguments we expect to find contributions of the form 
\eqref{eq:PolarFreeGF} where $\theta_{i}$ is replaced by angles reflecting
image contributions.  Here the second square root represents a distance and
is thus real and positive.  In order to satisfy the boundary conditions at
infinity one then chooses a branch of the first square root to satisfy $\Im
\sqrt{z} > 0$.  This has a branch cut along the positive real axis.

The ``distance'' square root can be written as 
$\sqrt{2r_{i}r}\lambda(\omega)$.  In the notation of 
section~\ref{sec:Kernel}, the argument of the square root is 
essentially $1-u$.  Using our knowledge of $\omega\leftrightarrow u$ 
we can determine the analytic properties in the $\omega$ plane 
corresponding to different choices of branch.

As our studies are limited to real energies our first choice was the 
branch $\Im \sqrt{z} > 0$ so that $G_0(r,r_{i};\omega)$ 
decays exponentially in all directions as $|\omega| \to
\infty$. Although this gives more choice for the contour $C$, it
creates a branch cut in the $\omega$--plane starting at $i\alpha$,
going towards the origin along the imaginary axis, along the positive
real axis until $\pi$ and then up the line $\pi + it$, $0 < t < \infty$.
The other branch cuts are obtained by suitable reflection and 
translation.
  
The image contributions, however, come from poles on the real axis so 
that this choice severly restricts (or at least complicates) the 
various contour deformations we need.  To avoid this we choosing the 
standard square root, which has a branch cut along the negative real 
axis, to obtain a branch cut in the $\omega$ plane which starts at 
$i\alpha$ and goes to $i\infty$, again with others related by symmetry 
operations.  The are depicted in Figure~\ref{fig:WedgeContourOmega} 
together with shaded regions depicting where the integrand decays at 
infinity.  We determine the latter by noting that $\Im \lambda(\omega) 
> 0$ requires $1-u$ to be in the upper half plane.  From 
Figure~\ref{fig:u} one can see that this corresponds to regions 
$\framebox{3}$ and $\framebox{4}$ which we can trace back to the 
$\omega$ plane using Figure~\ref{fig:wtot:w}.

Substituting the ansatz into the polar form of the Helmholtz equation, 
exchanging differentiation and integration, using 
Eqn.~\ref{eq:PolarHelmholtz} to eliminate the $r$ derivatives and 
integrating by parts twice, one obtains
\begin{equation}
\begin{split}
	(\nabla^{2} 
	& + z) G_{W}(r,\theta) 
	= \frac{1}{r^{2}}\int_{C}\
		\left( \frac{\partial^{2}f}{d\theta^{2}} -
			   \frac{\partial^{2}f}{d\omega^{2}} 
		\right) G_0(r,r_{i};\omega)\,d\omega - 
	\\
	& 
	 \left[ f(\omega,\theta)
	 		\frac{\partial G_0}{d\omega}(r,r_{i};\omega) -
	 		G_0(r,r_{i};\omega)
	 		\frac{\partial f}{d\omega} (\omega,\theta)
	 \right]^{B}_{A} 		
\end{split}
\end{equation}
If $C$ is chosen so that the latter ``complex flux'' term vanishes we 
see that the problem is reduced to a complexified 1--dimensional wave 
equation for $f(\omega,\theta)$ whose general solution is elementary
\begin{equation}
	f(\omega,\theta) = M(\omega + \theta) + N(\omega - \theta)
\end{equation}
Because the variables $r$ and $\theta$ are separated,
the Dirichlet boundary conditions for $\theta = 0, \phi$ can be satisfied 
with the choice  
\begin{subequations}
\begin{gather}
	f(\omega,\theta) = M(\omega + \theta) - M(\omega - \theta)
	\\
	M(\omega)  =  M(\omega + 2\phi)
	\label{eq:Periodicity}
\end{gather}
\end{subequations}
The remaining function $M(z)$ and the contour $C$ must now be 
determined from the unit source condition.

The unit source condition is satisfied if 
\begin{equation}\label{eq:unitSource}
	G_{W}(r,\theta) = G_{0}(r,\theta) + G_{1}(r,\theta) 
\end{equation}
where $G_{1}(r,\theta)$ has no sources in the domain $W$.  
This can be arranged with a construction which is essentially the 
method of images.

If 
\begin{equation}
	\lim_{z \to -\theta_{i}} M(z) 
		= \frac{1}{2\pi i}\frac{1}{z + \theta_{i}} 
	\label{eq:PoleCondition}
\end{equation}
then $M(\omega - \theta)$ has a pole of residue $1$ at $\omega = 
\theta - \theta_{i}$.  \ref{eq:unitSource} is satisfied if $C$ can be 
deformed to pick up this pole while ensuring that the remaining 
contour integral remains sourceless in 
$W$.

In contemplating an analytic function $M(z)$ satisfying 
Eqns.~\ref{eq:Periodicity} and \ref{eq:PoleCondition} we are naturally 
lead to consider the function
\begin{equation}
	M(z) = \lim_{N\to\infty}
	\frac{1}{2\pi i} \sum_{n = -N}^{N}
			\frac{1}{z + \theta_{i} - 2 n \phi}
\end{equation}
It is possible to choose other functions by adding further poles in a 
periodic way.  Such extra poles will generate features in the final 
solution which either spoil the uniform convergence or don't satisfy 
the boundary conditions.  The principal 
value--type construction is needed to obtain convergence.  The reader 
should note the resulting contour integral as a simple example of a 
conditionally convergent semiclassical series.  This sum is a textbook 
example~(\cite[p.261]{HilleVol165}).
\begin{equation}
	 \lim_{N\to\infty}
	\sum_{n = -N}^{N} \frac{1}{z -  n}
	= 
	\pi \cot\pi z
\end{equation}

$M(\omega - \theta)$ generates images at $\theta_{i} - 2 n \phi$.  
$M(\omega + \theta)$ generates images at $- \theta_{i} + 2 n \phi$.  
Some of these poles, with suitable contours, could generate unwanted 
source points images in the domain W.  This can be prevented in a 
rather elegant manner by an appropriate choice of the contour $C$.

We now have sufficient information to uniquely determine $C$.  
Requirements of uniform convergence dictate that $C$ must not be near 
singularities for any value of $\theta_{i}$.  In particular the 
contour cannot cut the real axis.  The ``complex flux'' can then only 
vanish if the contour starts and ends at $\infty$ in the lightly 
shaded regions in Figure~\ref{fig:WedgeContourOmega}.  We need loops 
in both the upper and lower half plane so that the contour can be 
deformed to the ``direct'' pole at $\omega = \theta - \theta_{i}$ 
so that the unit source condition is satisfied.  Furthermore the 
deformation must not pick up the poles in $M(z)$ which correspond to 
unwanted source terms.  A loop starting at $\Im \omega =
\infty $ in $-2\pi < \Re \omega < -\pi$ and finishing at $\Im \omega = 
\infty $ in $3\pi < \Re \omega < 4\pi$, together with it's lower half 
plane analog, for instance, will pick up unphysical sources.  The only 
choice satisfying all these conditions is the Carslaw contour.

The wedge Green's Function is thus
\begin{equation}    \label{eq:FullSolution}
	G_{W}(r,\theta) = \frac{1}{2\phi}
	\int_{C} 
		\left[
			P(\theta_{i} - \theta;\omega) - 
			P(\theta_{i} + \theta;\omega)
		\right] 
		G_0(r,r_{i};\omega)\,d\omega
\end{equation}

where 
\begin{equation}
  P(\alpha;\omega) = \frac{e^{i\pi(\alpha + \omega)/\phi}}
  {e^{i\pi(\alpha + \omega)/\phi} - 1}
  = \frac{1}
  {1 - e^{-i\pi(\alpha + \omega)/\phi}}
\end{equation}

The solution must satisfy reciprocity i.e.  the solution should be 
symmetric on interchanging $(r,\theta)$ and $(r_{i},\theta_{i})$.  
This is not manifest but can be demonstrated by (i) isolating the part 
of the integral involving $P(\theta_{i} - \theta;\omega)$; (ii) 
deforming its contour C to one which is symmetric under $\omega \to 
-\omega$; (iii) letting $\omega \to -\omega$ ($C \to C,G_0 \to G_0, 
d\,\omega \to -d\,\omega$) and (iv) using the result 
$-P(\alpha;-\omega) = - 1 + 
P(-\alpha;\omega)$.

The wedge solution also gives us a deeper understanding of why the 
method of images is exact iff $\phi = \frac{\pi}{n},n=1,2,\ldots$.  
For these angles the integrand is $2\pi$ periodic.  The contour can 
then be deformed into two pieces which have opposite sense, related by 
a $2\pi$ translation, picking up image points along the way.  
Periodicity causes the integrals to cancel.
The classical problem is also integrable at these angles.

\subsection{Wedge Kernel}

The integral kernels we want can now be calculated by differentiation.
Any kernel ending on the lower edge can be expressed in terms of 
\begin{equation}
\begin{split}
	K(r_{f},0; r_{i},\theta_{i})
	& \equiv 
	-\nabla G_{W}(r_{f},0; r_{i},\theta_{i})
		\cdot
		\left(-\hat{\theta}
		\right)
	\\
	& = 
	\frac{1}{r_{f}}
	\left.
	\frac{\partial G_{W}}
	{\partial\theta}
	(r_{f},\theta; r_{i},\theta_{i})	
	\right|_{\theta = 0}
\end{split}
\end{equation}
Using
\begin{equation}
	\left.
		\frac{\partial}{\partial\alpha} P(\alpha;\omega)	
	\right|_{\alpha = \theta_{i}}
	=
	\frac{\partial}{\partial\omega} P(\theta_{i};\omega)	
\end{equation}
and integrating by parts, we get
\begin{equation} \label{WedgeEq:Kernel}
	K(r_{f},0; r_{i},\theta_{i})  
	= 
	\frac{1}{\phi r_{f}} \int_{C} P(\theta_{i},\omega) 
			 \frac{i}{4} 
			 H_{1}^{(1)}(\mu \lambda(\omega)) 
			 \mu \lambda'(\omega)\,d\omega
\end{equation}

The kernel ending on the upper edge is simply
\begin{eqnarray}
	K(r_{f},\phi; r_{i},\theta_{i}) & = & 
	-\nabla G_{W}(r_{f},\phi; r_{i},\theta_{i})
		\cdot\hat{\theta}
	\\
	& = &
	-
	\frac{1}{r_{f}}
	\left.
	\frac{\partial G_{W}}{\partial\theta}(r_{f},\theta; r_{i},\theta_{i})	
	\right|_{\theta = \phi}
\end{eqnarray}
We proceed as for the other kernel, except that we need to use 
periodicity and change variables $\omega \to -\omega$ in a manner 
similar to the reciprocity proof, to 
obtain

\begin{equation}
	K(r_{f},\phi; r_{i},\theta_{i})  
	= 
	\frac{1}{\phi r_{f}} \int_{C} P(\phi - \theta_{i},\omega) 
			 \frac{i}{4} 
			 H_{1}^{(1)}(\mu \lambda(\omega)) 
			 \mu \lambda'(\omega)\,d\omega
\end{equation}
This is identical \eqref{WedgeEq:Kernel} except that the angle is 
measured from the ray $\theta = \phi$ in a clockwise sense. This is in 
fact required by reflection symmetry

\end{document}